\documentclass[aps,prd,preprint,preprintnumbers,showpacs,nofootinbib]{revtex4-1}
\usepackage{graphicx}
\usepackage{amsmath}
\usepackage{amssymb}
\usepackage[section] {placeins}
\usepackage{slashed}
\usepackage{color}
\usepackage{soul}
\usepackage{appendix}
\usepackage{placeins}
\begin{document}

\title{Can femtoscopic correlation function shed light on the nature of the lightest charm axial mesons? }
\author{K.~P.~Khemchandani$^{1}$\footnote{kanchan.khemchandani@unifesp.br}}
\author{ Luciano M. Abreu$^{2,3}$\footnote{luciano.abreu@ufba.br}}
\author{A.~Mart\'inez~Torres$^{2}$\footnote{amartine@if.usp.br}}
\author{F.~S.~Navarra$^2$\footnote{navarra@if.usp.br}}
\preprint{}

 \affiliation{
$^1$ Universidade Federal de S\~ao Paulo, C.P. 01302-907, S\~ao Paulo, Brazil. \\
$^2$Universidade de Sao Paulo, Instituto de Fisica, C.P. 05389-970, S\~ao Paulo,  Brazil.\\
$^3$Instituto de F\'isica, Universidade Federal da Bahia, 40210-340, Salvador, BA, Brazil.
}

\date{\today}

\begin{abstract}
 We present a coupled channel treatment of meson-meson dynamics, for systems with spin-parity $1^+$, and determine the corresponding amplitudes by solving the Bethe-Salpeter equations, which lead to the generation of two axial resonances when mesons are considered as the degrees of freedom in the model. One of them is narrow and has properties in good agreement with those of $D_1(2420)$. The other pole is wider, but its real and imaginary parts do not match well with the mass and width of  $D_1(2430)$.  The situation improves when a bare quark-model state is included, indicating that the dynamics at the quark level as well as among hadrons can describe the two states simultaneously. Further, we discuss that there exists a divergence in the value of the $D^*\pi$ scattering length determined through data coming from lattice QCD calculations and from heavy ion collisions. Such different values can be accommodated in the model by making small changes in the parameters while producing two poles having properties compatible with the two lightest $D_1$ states. With these results, we proceed to calculate the correlation function for the $D^{*+(0)}\pi^{0(+)}$ system for different sizes of the source. We discuss which scenarios can be useful to shed some light on the issue.
\end{abstract}


\maketitle

\section{Introduction}
Femtoscopic correlation functions have emerged as a powerful alternative tool to understand the nature of hadrons in recent times (to see the articles published in the current year, for example, see Refs.~\cite{ALICE:2022uso,ALICE:2022mxo,ALICE:2023wjz} or see the review of Ref.~\cite{Fabbietti:2020bfg}), though the idea is not new~\cite{Lisa:2005dd}. In the present manuscript, we analyze if the same idea can be implemented to  better determine the nature of the two lightest $D_1$ mesons, $D_1(2420)$ and  $D_1(2430)$, for which the  average values of  mass and width, as given by the Particle Data Group (PDG)~\cite{PDG2022}, are:
\begin{align}\nonumber
D_1(2420):&~~~ M=2422.1\pm 0.6 \text{ MeV,    } \Gamma=31.3\pm1.9\text{ MeV.}\\
D_1(2430):&~~~ M=2412\pm 9 \text{ MeV,    } \Gamma=314\pm29\text{ MeV.}\label{pdgvalues}
\end{align}

The evidence for the existence of two $D_1$ states, with the aforementioned properties, comes from a fit made to the experimental data on the $D^*\pi$ invariant mass distribution~\cite{LHCb:2015tsv,Belle:2003nsh}, where the $S$- and $D$-wave amplitudes are associated with $D_1(2430)$ and $D_1(2420)$, respectively.
From the theory side, two  $c\bar d$, $c \bar u$ states were predicted, as expected from heavy quark symmetry, with the spin-parity $J^\pi=1^+$, in a quark model based on one-gluon-exchange-plus-linear-confinement potential~\cite{Godfrey:1985xj}. Such states were found to arise from the $1^1P_1$ and  $1^3P_1$ interactions, with masses 2440  and 2490 MeV, respectively,~\cite{Godfrey:1985xj}. An admixture of the states is associated with the physical mesons, with a mixing angle of $-41^0$~\cite{Godfrey:1985xj}  (values of  35.3$^0$ or $-54.7^0$ for the mixing angle have been suggested based on the heavy quark symmetry~\cite{Barnes:2002mu}). Though the values of the masses determined in different quark models are not very different from those listed in Eq.~(\ref{pdgvalues}), a difference in the width of the order of factor 10 cannot always be explained. Here, we must clarify that we are referring to models which try to explain the mass and width from the same dynamics. Indeed, arguments based on heavy quark symmetry have been presented by Manohar and Wise~\cite{Manohar:2000dt} to explain that the ratio of the widths of the $D_1$ states could be of the order of factor 10, once the masses of the states are assumed and the quantum numbers $1^1P_1$ and  $1^3P_1$ are attributed to $D_1(2430)$ and $D_1(2420)$, respectively (without considering a mixing between the two states).  Indeed, as shown in Refs.~\cite{Ferretti:2015rsa,Cahn:2003cw}, a consideration of a mixing of the two quark model states through spin-orbit interactions leads to a ratio of widths of the two states to be far less than that found in the experimental data. In yet another work~\cite{Zhou:2011sp}, it is shown that the ratio of the widths of the unmixed poles is around 2.5, which is not compatible with Eq.~(\ref{pdgvalues}). The same work shows, however, that a mixing of the two states through the consideration of hadron loops can describe the widths of the two $D_1$ states and suggests that such a finding is compatible with the mixing angle proposed in Ref.~\cite{Barnes:2002mu}.  A similar conclusion has been reached in Ref.~\cite{Ni:2021pce}. Further, a correction of about $-(5.7\pm2.9)^0$ is proposed in Ref.~\cite{Wang:2018psi}. In another study~\cite{Godfrey:2005ww}, predictions of radiative decay widths have been made to test the value of the mixing angle based on the heavy quark symmetry. Thus, it seems that different types of mixing of the $1^1P_1$ and  $1^3P_1$ states lead to similar masses but different values of the widths of the $D_1$ states and that the consideration of hadron loops or meson clouds is useful in better describing the properties of the lowest-lying $D_1$ states.  
 
On the other hand, the findings of models based purely on hadron dynamics show that although two low-lying $D_1$ states are also always found,   their properties do not coincide with those found in experiments.  Besides the difference in the properties of the states found in different models in comparison with those found experimentally, as we discuss below, there exists a deviation in the scattering length of $D^*\pi$ determined from data on lattice QCD~\cite{Mohler:2012na}  and from heavy ion collisions~\cite{ALICE:2024bhk}. Indeed information on observables related to the $D^*\pi$ channel is very relevant to understanding the nature of the $D_1$ states under discussion since it is the main decay channel for such states.  Given the situation, one of the purposes of this manuscript is to propose a model that can describe the mass and width of the two lowest-lying $D_1$ states.  Our attempt shows that an interplay of quark-hadron degrees of freedom can be useful in describing the aforementioned states (in line with the suggestions made in Ref.~\cite{Zhou:2011sp}).   Yet another aim is to investigate if femtoscopic correlation functions related to channels and source sizes different from those considered in Ref.~\cite{ALICE:2024bhk} can be useful in resolving the situation. We 
 show that the extraction of data from smaller source sizes and for channels dominated by strong interactions (and not needing Coulomb interactions) can be used to settle the value of the $D^*\pi$ scattering length, which can contribute to a better understanding of the properties of the two lowest-lying $D_1$ states. 

\subsection{A short summary of model findings.}
Let us summarize the results of some of the works attempting to describe the dynamical origin of the mass and width of the lightest axial states with charm. For example, it was pointed out in Ref.~\cite{Du:2017zvv} that all the low-lying positive-parity heavy open-flavor mesons can be understood as hadronic molecules. Such a claim is motivated by facts like the masses of the lightest open-charm mesons with strangeness are lower than their nonstrange counterparts. In Ref.~\cite{Du:2017zvv}, scattering equations were solved using kernels based on the unitarized chiral perturbation theory for heavy mesons and by fixing the free parameters to fit the scattering lengths determined from lattice QCD calculations. Within such a framework, two poles were found for $D_1$ mesons, with the mass and the  half-width being $2247^{+5}_{-6}, ~107^{+11}_{-10}$ and $2555^{+47}_{-30}, ~203^{+8}_{-9}$, where the lower one was associated with $D_1(2430)$. It is argued in the former work that the mass value, obtained from a  Breit-Wigner fit, given by  PDG for $D_1(2430)$ needs to be revised. A narrow pole, to be associated with $D_1(2420)$, however, is not found. Curiously, a very different quark-model calculation~\cite{Abreu:2019adi} obtained mass values for the $D_1$ states which are very similar to those of Ref.~\cite{Du:2017zvv}. 
Information on $D_1$ mesons is also available from $^{2S+1}l_J={}^3S_1,~{}^3D_1$ amplitudes determined for the $D^* \pi$ system with lattice QCD calculations~\cite{Lang:2022elg}. A broad and a narrow state is found, respectively, in the former and the latter amplitudes. It is worth mentioning that the broad $^3S_1$ state, which is related to $D_1(2430)$, has been found to appear at mass 2397 MeV even though the pion mass is larger than the corresponding physical mass, indicating that the physical pole mass could be lower. Updated calculations, with contributions from channels like $D\pi\pi$, are expected to be determined in future~\cite{Lang:2022elg}. As we will discuss, in fact adding coupled channels like $D\rho$ can be important to better understand the properties of the $D_1$ states.
 
It is worth mentioning that several other attempts have been made to simultaneously describe the properties of the $D_1$ states~\cite{DiPierro:2001dwf,Colangelo:2004vu,Kolomeitsev:2003ac,Guo:2006rp,Gamermann:2007fi,Malabarba:2022pdo,Coito:2011qn,Burns:2014zfa}, claiming different attributions to their dynamical origin
 
 \subsection{Our idea}
The main purpose of our work is to determine the femtoscopic correlation functions for the $D^{*+(0)}\pi^{0(+)}$ systems, which are free from Coulomb interactions. These systems are different from those investigated in Ref.~\cite{ALICE:2024bhk}. The idea is also to calculate the correlation function for different sizes of the source and study if there exist more ideal conditions to determine the nature of the $D^*\pi$ strong interaction.  To do this we need the amplitudes of the $D^*\pi$ system which carry the information on the $D_1$ states. We proceed by extending the framework developed in Refs.~\cite{Gamermann:2007fi,Malabarba:2022pdo} by adding a quark-model pole to the lowest order $D^*\pi$ amplitude.   As shown in Ref.~\cite{Malabarba:2022pdo} the dynamics of $D^*\pi$ and coupled channels leads to the formation of a state whose mass and width are in excellent agreement with those of $D_1(2420)$.  It is worth emphasizing that the aforementioned pole couples strongly to $D\rho$ but very weakly to  $D^*\pi$, which naturally explains its narrow width that is in contrast with the one of $D_1(2430)$. Next, we should add that yet another pole, coupling mostly to $D^*\pi$, appears in the complex plane in the formalism of Ref.~\cite{Malabarba:2022pdo}, around $2222-i61$ MeV~\footnote{The position of the corresponding pole in Ref.~\cite{Gamermann:2007fi} was $2311.24-i115.68$ MeV, which is in better agreement with the properties of $D_1(2430)$. However, recall that the narrow pole found in  Ref.~\cite{Gamermann:2007fi} appears at $\sim$2526 MeV which is far from the mass of $D_1(2420)$.}. This latter pole is not in good agreement with the properties of $D_1(2430)$. To improve this accordance, we add a bare quark-model pole to the lowest order $D^*\pi$ amplitude of Ref.~\cite{Malabarba:2022pdo}.

On solving Bethe-Salpeter equations with such kernels, a broad state, as well as a narrow state, are found in the resulting amplitudes. 
The narrow state related to $D_1(2420)$ is as found in Ref.~\cite{Malabarba:2022pdo}. The broad state has properties closer to those of $D_1(2430)$.  
To further compare the results of the model with the known data we calculate the  $D^*\pi$ scattering length.  We recall that scattering length values for the $D\pi$ and  $D^*\pi$ channels are determined in Ref.~\cite{Mohler:2012na} within the lattice QCD framework, using a pion mass of 266 MeV. Also, in Refs.~\cite{Liu:2012zya,Guo:2018tjx}, $D\pi$ and other systems are investigated in the lattice QCD formulation, the results of which are used to fix the parameters of the next-to-leading order term of the chiral Lagrangian and the $D\pi$ scattering length is obtained with the physical pion mass. The resulting values are found to agree with those obtained within unitarized chiral perturbation theory and heavy quark symmetry~\cite{Guo:2009ct,Abreu:2011ic,Geng:2010vw}. Values of the $D\pi$ and $D^*\pi$ scattering lengths have also been determined by the ALICE collaboration~\cite{ALICE:2024bhk} (see also Refs.~\cite{Alice,Alice2,Alice3,Alice4}), though, enigmatically, the value determined seems to be different to the results of all the aforementioned works~\cite{Liu:2012zya,Guo:2018tjx,Guo:2009ct,Abreu:2011ic,Geng:2010vw}. The question that arises is if the $D\pi$ and $D^*\pi$ scattering lengths should be as obtained in Refs.~\cite{Liu:2012zya,Guo:2018tjx,Guo:2009ct,Abreu:2011ic} or as in Ref.~\cite{ALICE:2024bhk}. It is the purpose of our paper to investigate if femtoscopic studies of systems or source sizes different from those considered in Ref.~\cite{ALICE:2024bhk} can be used to resolve this matter. Hence we contemplate two possibilities in our model which can produce different scattering lengths and calculate the correlation function for $D^{*0(+)}\pi^{+(0)}$, for different values of the source size. We present results for different source sizes and discuss if such a tool can be used to resolve the issue and discuss the implications of the two scenarios on the properties of the two lowest-lying $D_1$ mesons.

The article is organized as follows. We first discuss the formalism used to determine the amplitudes for different meson-meson systems and present the convention followed to evaluate the scattering length. We give a summary of the scattering lengths obtained within different works, which leads to the consideration of two different parametrizations. We then show the amplitudes, discuss the properties of the resulting poles, and compare the scattering lengths obtained with other works. In a subsequent section, we present the details of the calculations of the correlation function. Finally, we summarize the conclusions of the work.
 
 \section{Determining the scattering amplitudes}
 \subsection{Formalism to solve the scattering equations}
 In the present work, we start by following the formalism of Refs.~\cite{Gamermann:2007fi,Malabarba:2022pdo}, where nonperturbative interactions between vector and pseudoscalar mesons are considered on the basis of a broken SU(4) symmetry, and add a bare quark-model pole in the kernel. To be more explicit, we begin by considering the lowest order amplitude~\cite{Gamermann:2007fi,Malabarba:2022pdo},
 \begin{align}
 t_{ij}=\frac{C_{ij}}{4f^2}\left(s-u\right) \vec \epsilon\cdot\vec \epsilon^\prime,\label{tamp}
 \end{align}
where $f$ is the pion decay constant, taken to be 93 MeV, $s$, $u$ are Mandelstam variables, $\epsilon \left(\epsilon^\prime\right)$ represents the polarization (three-) vector for the incoming (outgoing) vector meson, and $C_{ij}$ are constants for different $i,~j$ (initial, final state). 

Though the amplitude in Eq.~(\ref{tamp}) is obtained in Ref.~\cite{Gamermann:2007fi} by using the SU(4) symmetry, the same can also be determined, as we do, by considering the following Lagrangians~\cite{Bando:1984ej,Bando:1987br,Meissner:1987ge,Harada:2003jx}
\begin{align}\nonumber
\mathcal{L}_{PPV}&=-ig_{PPV}\langle V^\mu\left[P,~\partial_\mu P\right]\rangle,\\
\mathcal{L}_{VVP}&=\frac{g_{VVP}}{\sqrt{2}}\epsilon^{\mu\nu\alpha\beta}\langle\partial_\mu V_\nu\partial_\alpha V_\beta P\rangle
 \end{align}
to calculate the contribution of a vector-meson exchange in the $t$-channel and by applying the approximation $t\to 0$ at energies near the threshold. We write the  SU(4) matrix for pseudoscalar ($P$) and vector ($V$) fields as
 \begin{align}\nonumber
 P&=\left(\begin{array}{cccc}\frac{\eta}{\sqrt{3}}+\frac{\eta^\prime}{\sqrt{6}}+\frac{\pi^0}{\sqrt{2}}&\pi^+&K^+&\bar D^0\\ \pi^-&\frac{\eta}{\sqrt{3}}+\frac{\eta^\prime}{\sqrt{6}}-\frac{\pi^0}{\sqrt{2}}&K^0&D^-\\K^-&\bar K^0&-\frac{\eta}{\sqrt{3}}+\sqrt{\frac{2}{3}}\eta^\prime&D^-_s\\D^0&D^+&D_s^+&\eta_c\end{array}\right),\\\nonumber
  V_\mu&=\left(\begin{array}{cccc}\frac{\omega}{\sqrt{2}}+\frac{\rho^0}{\sqrt{2}}&\rho^+&K^{*+}&\bar D^{*0}\\ \rho^-&\frac{\omega}{\sqrt{2}}-\frac{\rho^0}{\sqrt{2}}&K^{*0}&D^{*-}\\K^{*-}&\bar K^{*0}&\phi&D^{*-}_s\\D^{*0}&D^{*+}&D_s^{*+}&J/\Psi\end{array}\right),
 \end{align}
which are slightly different from those in Refs.~\cite{Gamermann:2007fi,Malabarba:2022pdo}.  The difference arises from including the mixing between $\eta-\eta^\prime-\eta_c$ and $\omega-\phi-J/\Psi$, which gives slightly different amplitudes.  We, thus, provide the values of $C_{ij}$ in Table~\ref{Table:cij}.

\begin{table}[ht!]
\caption{Values of the $C_{ij}$ constants appearing in Eq.~(\ref{tamp}) for different processes in the isospin 1/2 configuration. Here $\gamma=\left(\frac{m_L}{m_H}\right)^2$, where $m_L=800$ MeV and $m_H=2050$ MeV are average masses of the light and heavy vector mesons, as defined in Refs.~\cite{Gamermann:2007fi}. In the case of isospin 3/2, we have two channels: $D^*\pi$ and $D\rho$ with the diagonal values of $C_{ij}$ being 1 and the nondiagonal being $\gamma$. 
}\label{Table:cij}
\begin{ruledtabular}
\begin{tabular}{c|cccccccccc}
                 & $\pi D^*$ & $D\rho$       & $\bar KD_s^*$        & $D_s\bar K^*$  & $\eta D^*$ & $D\omega$                      & $\eta_c D^*$ & $D J\Psi$             & $D\phi$ &$\eta^\prime D^*$\\\hline
 $\pi D^*$ & $-2$         &$\gamma/2$&  $-\sqrt{\frac{3}{2}}$&0                          &0                &$\frac{\sqrt{3}\gamma}{2}$&0                    &$-\sqrt{\frac{3}{2}}\gamma$&0&0\\
 $D\rho$   &                 & $-2$            & 0                             &$\sqrt{\frac{3}{2}}$ & $-\sqrt{\frac{1}{2}}\gamma$    &0 & $\sqrt{\frac{3}{2}}\gamma$ &0&0& $-\frac{\gamma}{2}$\\
 $\bar KD_s^*$   &&&$-1$&0&$-\frac{2}{\sqrt{3}}$&0&0&$-\gamma$&$\gamma$&$\frac{1}{\sqrt{6}}$\\
$D_s\bar K^*$ & &&& $-1$ & $-\frac{\gamma}{\sqrt{3}}$ & $-\frac{1}{\sqrt{2}}$&$-\gamma$&0&1&$\sqrt{\frac{2}{3}}\gamma$\\
 $\eta D^*$&&&& &0&$\frac{\gamma}{\sqrt{6}}$&0&$-\frac{\gamma}{\sqrt{3}}$&0&0\\
 $D\omega$     &&&& &&0&$-\frac{\gamma}{\sqrt{2}}$&0&0&$\frac{\gamma}{2\sqrt{3}}$\\
 $\eta_cD^*$ &&&& &&&0&$\gamma$&0&0\\
 $D J\Psi$&&&& &&&&0&0&$-\frac{\gamma}{\sqrt{6}}$\\
 $D\phi$  &&&& &&&&&0&0\\
 $\eta^\prime D^*$&&&& &&&&&&0
\end{tabular}
\end{ruledtabular}
\end{table}

Besides Eq.~(\ref{tamp}), we consider contributions coming from a pseudoscalar exchange through box diagrams  as obtained in Ref.~\cite{Malabarba:2022pdo} for the $D\rho \to D^*\pi \to D\rho$.  If we solve the Bethe-Salpeter equation 
\begin{align}
T=V+VGT,\label{BSE}
\end{align}
with such amplitudes, just as done in Ref.~\cite{Malabarba:2022pdo}, we find results very similar to those of Ref.~\cite{Malabarba:2022pdo}. Thus, as expected, the mixing between $\eta-\eta^\prime-\eta_c$ and $\omega-\phi-J/\Psi$ do not give important contributions. 

\begin{figure}[h!]
    \centering
    \includegraphics[width=0.4\textwidth]{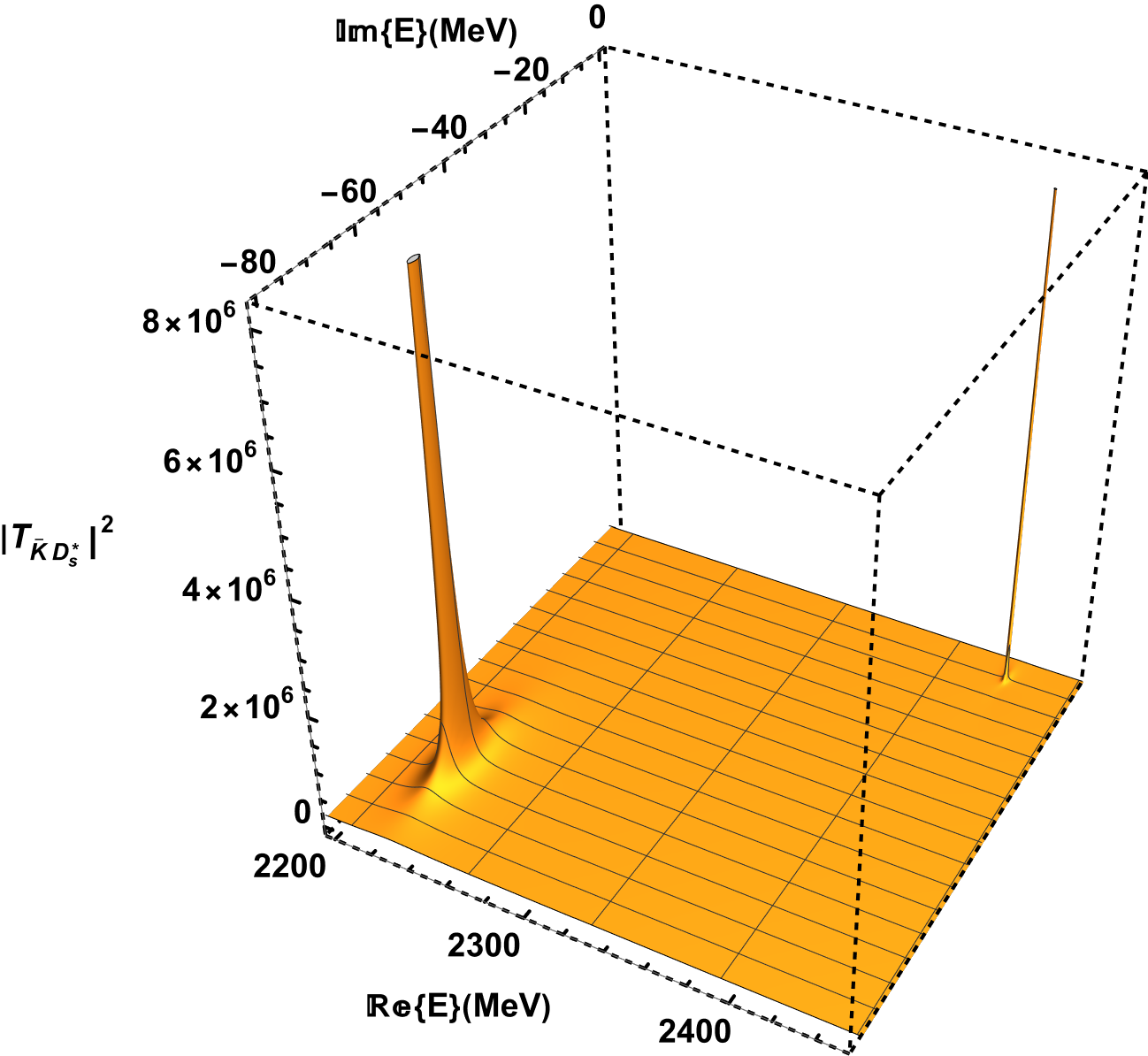}
    \caption{Two poles appearing in the $\bar K D_s^*$ amplitude. While the narrower pole can be associated with $D_1(2420)$, the broader pole does not represent the properties of $D_1(2430)$. It should be mentioned that here we have chosen to show the $\bar K D_s^*$ amplitude as an example. The poles are seen in all the coupled channels, as the case should be.  }
    \label{twopoleOld}
\end{figure}

The loop function, $G$, in  Eq.~(\ref{BSE}), for the $k$th channel, is~\cite{Gamermann:2007fi,Malabarba:2022pdo}
\begin{align}\nonumber
&G_k\left(\sqrt{s}\right)=\int\frac{d^4q}{\left(2\pi\right)^4}\frac{1}{\left(P-q\right)^2-M^2_k}\frac{1}{q^2-m_k^2}\\\nonumber
&=\frac{1}{16\pi^2}\Biggl[a_k\left(\mu\right)+\ln{\frac{m_k^2}{\mu^2}}+\frac{M_k^2-m_k^2+s}{2s}\ln{\frac{M_k^2}{m_k^2}}+\frac{q_k}{\sqrt{s}}\Biggl(\ln{\frac{s-\left(M_k^2-m_k^2\right)+2q_k\sqrt{s}}{-s+\left(M_k^2-m_k^2\right)+2q_k\sqrt{s}}}\Biggr.\Biggr.\\
&+\Biggl.\Biggl.\ln{\frac{s+\left(M_k^2-m_k^2\right)+2q_k\sqrt{s}}{-s-\left(M_k^2-m_k^2\right)+2q_k\sqrt{s}}}\Biggr)\Biggr]\left(1+\frac{q_k^2}{3M_k^2}\right),\label{gfn}
\end{align}
with $P$ representing the total four-momentum of the system and $q$ the four-momentum of one of the mesons, $M_k$ ($m_k$) standing for the mass of the vector (pseudoscalar) meson in the $k$th channel, $a_k$ and $\mu$ denoting a subtraction constant needed to regularize the divergent nature of the $G$-function and the regularization scale, respectively.
The values for the regularization parameters are taken to be~\cite{Malabarba:2022pdo} $\mu=1500$, $a=-1.45$, and we consider the convolution of  $G$-function over the finite widths of $\rho$ and $K^*$. The resulting amplitudes show the presence of a state with mass $~\sim 2428$~MeV and width $\sim 33$~MeV on the real axis, as in Ref.~\cite{Malabarba:2022pdo}.  This latter state couples strongly to the $D\rho$ channel and weakly to the $D^*\pi$ channel,  and its properties are in excellent agreement with those of $D_1(2420)$. There appears another pole at $2220-i61$~MeV, which couples mostly to $D^*\pi$ but which cannot be related to $D_1(2430)$ (see Fig.~\ref{twopoleOld}).

\begin{figure}[h!]
    \centering
    \includegraphics[width=0.4\textwidth]{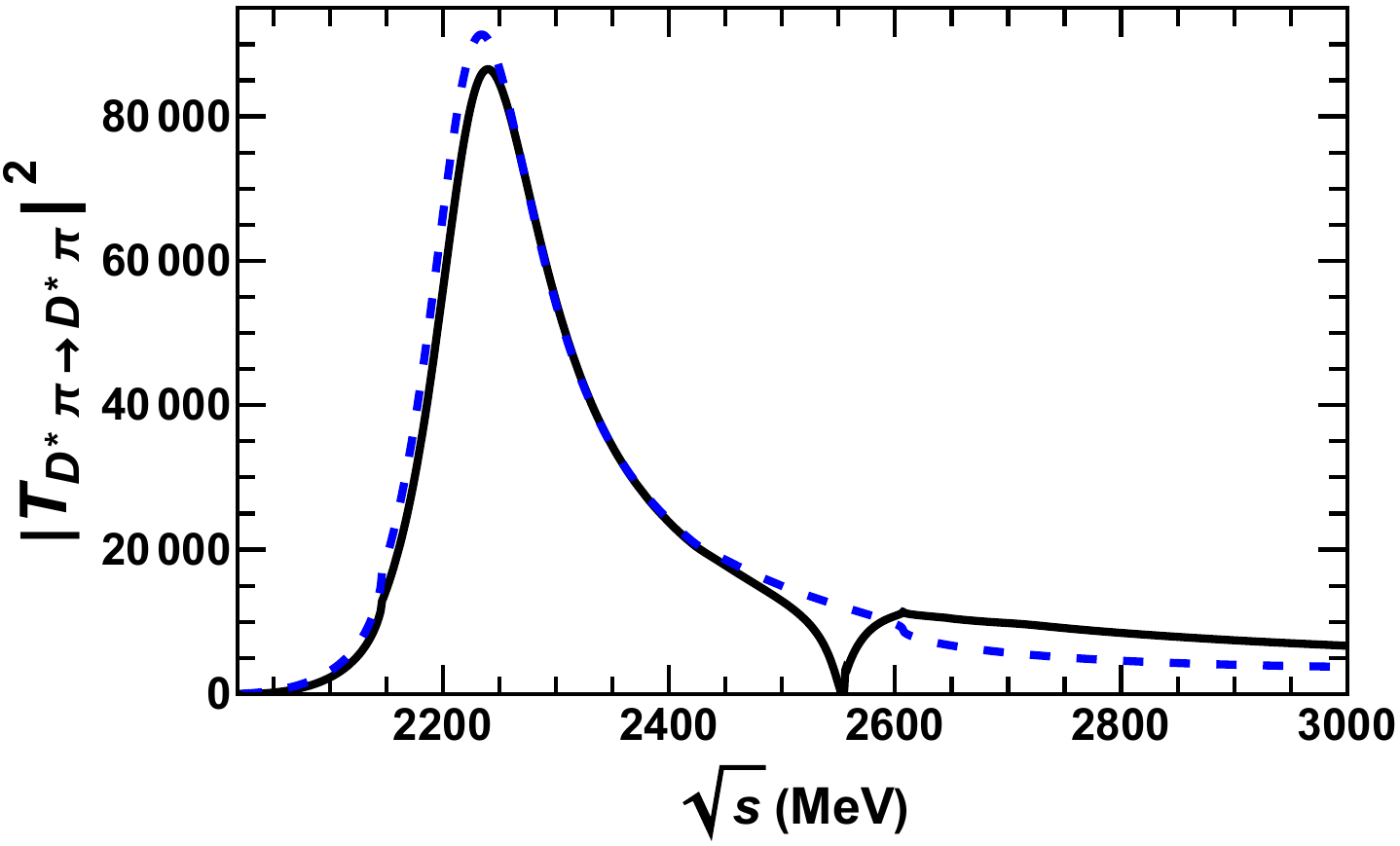}~~~~\includegraphics[width=0.4\textwidth]{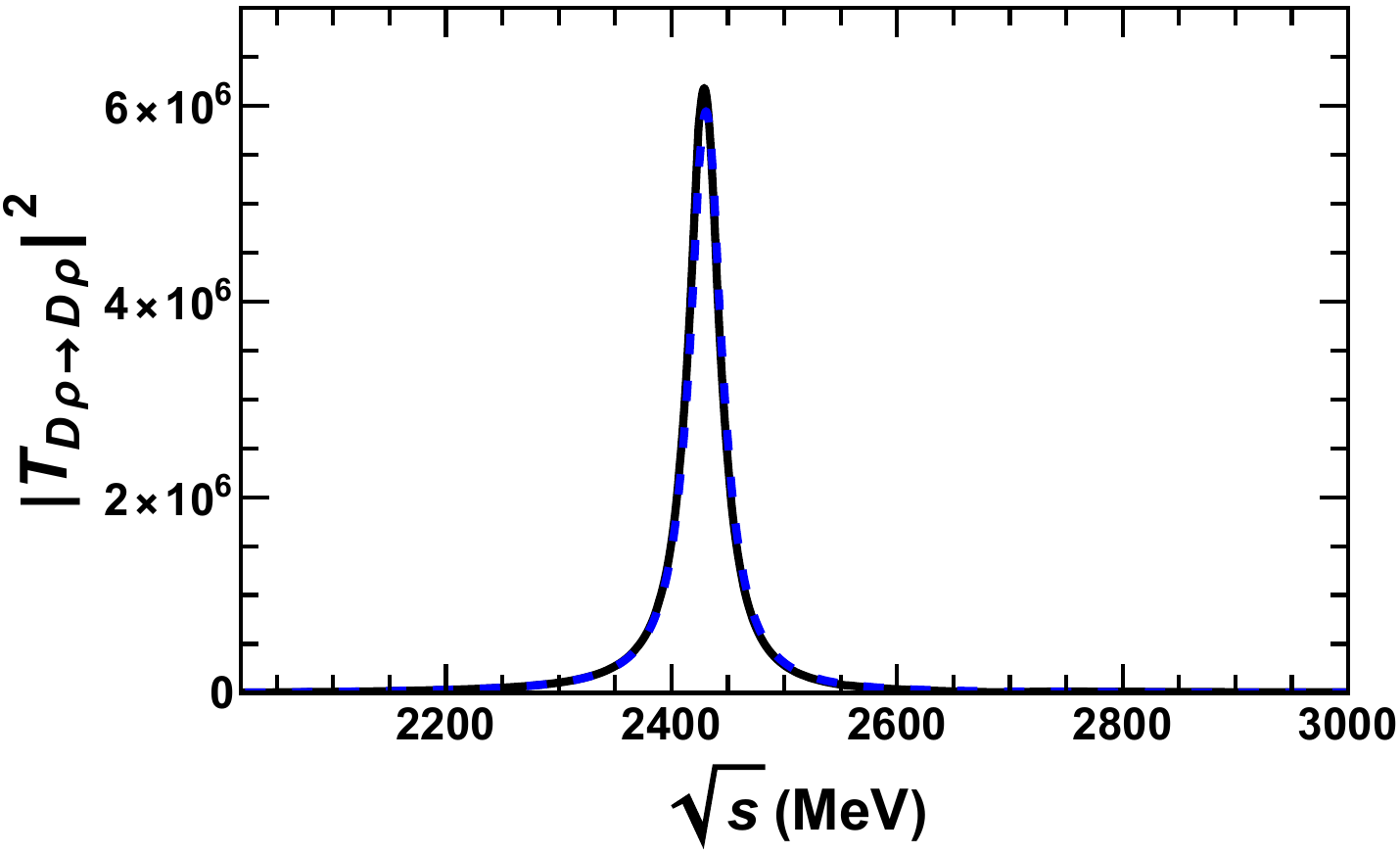}
    \caption{Projection of the poles shown in Fig.~\ref{twopoleOld} on the real axis, shown by depicting the squared amplitudes for the $D^*\pi$ and $D\rho$ channels. The solid lines are the results of solving the scattering equation with ten coupled channels (given in Table~\ref{Table:cij}), while the dashed lines result from considering only the first four channels, which are $D^*\pi$, $D\rho$, $\bar K D^*_s$, and $D_s \bar K^*$. The two results look very similar, except for a $\eta D^*$ cusp effect seen around 2556 MeV in the $D^*\pi$ amplitude, shown as a solid line. }
    \label{2poleRealaxes}
\end{figure}

 Notice that the half-width of the narrow pole is around 5 MeV, in the complex plane, but the width on the real axis becomes 33 MeV on consideration of the finite widths of the $\rho$ and $K^*$ mesons. The width of the lower energy pole also increases to ~130 MeV, but it still remains too small for the purpose of its association with $D_1(2430)$. Besides the mass also remains too low.  We show the $D^*\pi$ and $D\rho$ squared amplitudes, on the real axis, as solid lines in Fig.~\ref{2poleRealaxes}. 
The results shown as dashed lines, Fig.~\ref{2poleRealaxes}, correspond to those obtained by considering only the first four channels of Table~\ref{Table:cij}. It can be seen the results are almost unchanged, indicating that the most relevant channels to study the mentioned $D_1$ states are $D^*\pi$, $D\rho$, $\bar K D^*_s$, and $D_s \bar K^*$. The noncompatibility between the properties of $D_1(2430)$ and the lower energy pole seen in Fig.~\ref{twopoleOld}, whose effect on the real axis is shown in the left panel of Fig.~\ref{2poleRealaxes}, shows that something is missing in the model.
We must recall at this point that the degrees of freedom considered in our model, so far, are hadrons and other inputs could be necessary to better describe the properties of $D_1(2420)$ and $D_1(2430)$ simultaneously.
 
\begin{figure}[h!]
    \centering
    \includegraphics[width=0.5\textwidth]{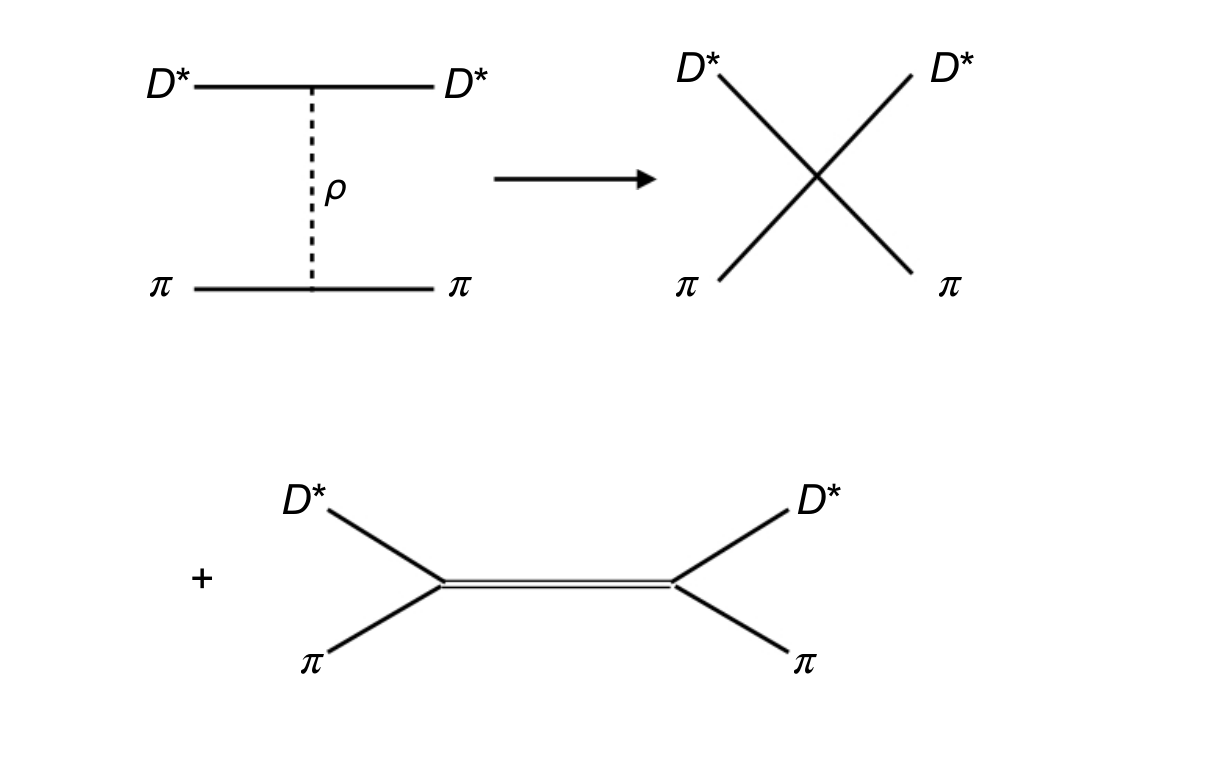}
    \caption{Lowest order amplitude for the $D^*\pi$ channel. The upper diagram originates from a vector exchange in the $t$-channel, whereas the lower one arises from the exchange of a bare quark-model pole in the $s$-channel. }
    \label{kernel}
\end{figure}

For this purpose, as a next step, we try adding a bare quark-model pole to the lowest order amplitude for the $D^*\pi$ channel, following Refs.~\cite{Nieves:2019nol,Cincioglu:2016fkm,Albaladejo:2016ztm,Albaladejo:2018mhb,Geng:2006yb,MartinezTorres:2014kpc}, as\footnote{The potential in Eq. (6) should be considered as an effective
interaction added to one of the channels mimicking the effect of adding a term to each
channel. Such an ansatz is followed to minimize the number of free parameters}
\begin{align}
V_{QM}=\pm\frac{g^2_{QM}}{s-M_{QM}^2},\label{vqm}
\end{align}
where the mass, $M_{QM}$, can be taken from different quark model calculations~\cite{Godfrey:1985xj,Ferretti:2015rsa,Abreu:2019adi} and $g_{QM}$ can be adjusted to obtain a better agreement between the lower energy pole shown in Fig.~\ref{twopoleOld} and the properties of $D_1(2430)$. The resulting lowest order amplitude for $D^*\pi$ is then the sum of diagrams shown in Fig.~\ref{kernel}. 

Before proceeding further we must clarify why we add one quark-model pole when two of them appear in Refs.~\cite{Godfrey:1985xj,Ferretti:2015rsa,Abreu:2019adi}.  The reason is that one of the states generated by meson-meson dynamics, in our formalism (which is based on Refs.~\cite{Gamermann:2007fi,Malabarba:2022pdo}), is in very good agreement with the properties of $D_1(2420)$. The implication of such a finding is that the contribution of the quark-model component to the $D_1(2420)$ wave function must be small, such as to keep the properties of the pole related to $D_1(2420)$ mostly unchanged.  It should also be mentioned that the regularization parameter used to solve the Bethe-Salpeter equation can embed contributions from other types of dynamics not explicitly considered in the model (see discussions in Refs.~\cite{Hyodo:2011qc,MartinezTorres:2011pr,Aceti:2012dd}).

 \subsection{Scattering lengths and related uncertainties}\label{scatlen}
Before depicting the amplitudes for different channels, we digress a bit towards the discussion of another observable, on which information is available from different sources. The observable being referred to is the scattering length, which we calculate for different channels through the relation
\begin{align}
  a_i=\frac{T_{ii}}{8\pi \sqrt{s}}.  \label{alength}
\end{align}
We compare the results for the $D^*\pi$ channel with the information available from lattice QCD calculations~\cite{Mohler:2012na}.  The former work determined the isospin 1/2 scattering lengths for $D\pi$ and $D^*\pi$ channels and obtained the following values:
\begin{align}\nonumber
 a^{(1/2)}_{D\pi}&=0.81\pm 0.14 \text{ fm,}\\
 a^{(1/2)}_{D^*\pi}&=0.81\pm 0.17\text{ fm,}\label{alat0}
\end{align}
at a pion mass of 266 MeV. As can be seen, the values for the two channels are very similar. Such a finding can be understood by invoking arguments of heavy quark symmetry. It is important to stress here that a sign convention opposite to that given in Eq.~(\ref{alength}) is followed in Ref.~\cite{Mohler:2012na}. Further, scattering lengths for systems like $D\bar K$, $D_s\pi$, $D_s K$, and $D\pi$ in isospin 3/2 configurations have been obtained on a lattice in Ref.~\cite{Liu:2012zya}. Using the values of the low energy constants of the chiral Lagrangian fixed from such a study, the scattering length for $D\pi$ in isospin 1/2 is determined. The results from the former works, obtained at the physical pion mass, can be summarized as
\begin{align}\nonumber
 a^{(1/2)}_{D\pi}&=0.37\pm0.01\text{ fm,}\\
 a^{(3/2)}_{D\pi}&=-0.100(1)\text{ fm.}\label{alat}
\end{align}
It is worth mentioning here that the value of $a^{(1/2)}_{D\pi}$ at $m_\pi=266$~MeV is obtained to be $2.30^{+2.40}_{-0.66}$ fm in Ref.~\cite{Liu:2012zya}, which is higher than that of Ref.~\cite{Mohler:2012na} [given in Eq.~(\ref{alat0})].
On the other hand, results compatible with Eq.~(\ref{alat}) have been obtained in Refs.~\cite{Guo:2018tjx,Guo:2009ct}, by using effective theories based on chiral and heavy quark symmetry and by constraining the values of the unknown parameters to fit the results available from lattice QCD. Somewhat smaller values for the isospin 1/2, around 0.2 fm, are found in Refs.~\cite{Abreu:2011ic,Geng:2010vw}, which are compatible with the results obtained using the leading order term of the chiral effective Lagrangian~\cite{Guo:2018tjx,Guo:2009ct}. 

Recent results on the $D^*\pi$ scattering length are available from an alternative source, from the Alice collaboration~\cite{ALICE:2024bhk}, which seem to be in disagreement with the aforementioned values. Let us denote the values given in Refs.~\cite{ALICE:2024bhk} as $\tilde a^{(I)}$. Using such a notation, the values in  Refs.~\cite{ALICE:2024bhk} can be summarized as
\begin{align}\nonumber
 \tilde a^{(1/2)}_{D\pi}&=0.02\pm0.03\pm0.01 \text{ fm}, \\\nonumber
 \tilde a^{(3/2)}_{D\pi}&=0.01\pm 0.02\pm 0.01 \text{ fm}, \\\nonumber
 \tilde a^{(1/2)}_{D^*\pi}&=-0.03\pm0.05\pm0.02 \text{ fm},\\
\tilde a^{(3/2)}_{D^*\pi}&=0.05\pm0.04\pm0.02 \text{ fm}.\label{aAlice}
\end{align}

In such a scenario, with different values of scattering lengths obtained from different sources, with one relying on lattice QCD calculations and the other being the experimental data obtained by the Alice collaboration, we find it instructive to investigate if femtoscopic correlation functions measured for source sizes different to the one in Ref.~\cite{ALICE:2024bhk} can resolve this puzzle. We will eventually show that the correlation functions of $D^{*0}\pi^+$ and $D^{*+}\pi^0$ are sensitive to the different values of the scattering lengths, and the difference is more marked for smaller source sizes. We hope that our findings can motivate the determination of data satisfying such conditions.

\subsection{Amplitudes: Results and discussions}
As mentioned in the preceding discussions, we add Eq.~(\ref{vqm}) to the amplitudes already considered in Refs.~\cite{Gamermann:2007fi,Malabarba:2022pdo} with the idea of having the presence of a wide as well as a narrow $D_1$ state around 2430 MeV. In addition for the amplitudes to show states with properties compatible with those of $D_1(2420)$ and $D_1(2430)$, we require the values of the $D^*\pi$ scattering lengths to be in agreement with Eqs.~(\ref{alat}) or (\ref{aAlice}). For this purpose, we shall consider two different sets of values for the parameters $M_{QM}$ and $g_{QM}$, and refer to the cases as models A and B. It must be clarified here that there is little room for changing the only other possible parameter of the model, which is the subtraction constant used to regularize the loop function since the properties of $D_1(2420)$ are well described by our amplitudes as discussed in Ref.~\cite{Malabarba:2022pdo}. Note that the scale $\mu$ and the subtraction constant $a_k$ appearing in Eq.~(\ref{gfn}) are not two parameters. Both are related to each other and, hence, together they account for one parameter only.  

It should be emphasized that adding the bare pole to the $D^*\pi$ amplitude does not affect the narrow pole associated with $D_1(2420)$ and only the wider pole changes its position in the complex plane. The ansatz followed in the work, as summarized in Fig~\ref{kernel}, is chosen on purpose since a good description of $D_1(2420)$ is already obtained with the meson degrees of freedom. Thus, we end up finding two poles which can be related to $D_1(2420)$ and $D_1(2430)$. The added quark-model pole, on the other hand, in the two models discussed in the following sections, becomes a virtual pole~\footnote{A pole which appears below the threshold of a given channel, on the unphysical Riemann sheet, is called as a virtual pole.}.  

\subsubsection{Model A}\label{modelA}
One of the choices we make is to write Eq.~(\ref{vqm}) as
\begin{align}
V_{QM}=-\frac{6000^2}{s-2440^2},\label{vqmA}
\end{align}
where the value of the mass, $M_{QM}=2440$~MeV, is taken from the quark model of Ref.~\cite{Godfrey:1985xj}. Such a choice, together with $g_{QM}=6000$~MeV, leads the lower energy pole in Fig.~\ref{twopoleOld} to move to $E-i\Gamma/2=2268-i100$ MeV.  We stress that we solve the Bethe-Salpeter equation by considering the first four channels shown in Table~\ref{Table:cij}. As already shown in Fig.~\ref{2poleRealaxes}, the first four channels of Table~\ref{Table:cij} are found to be the most relevant ones for studying $D_1$ states in the energy region of $2150-3000$ MeV. 

\begin{figure}[h!]
    \centering
    \includegraphics[width=0.45\textwidth]{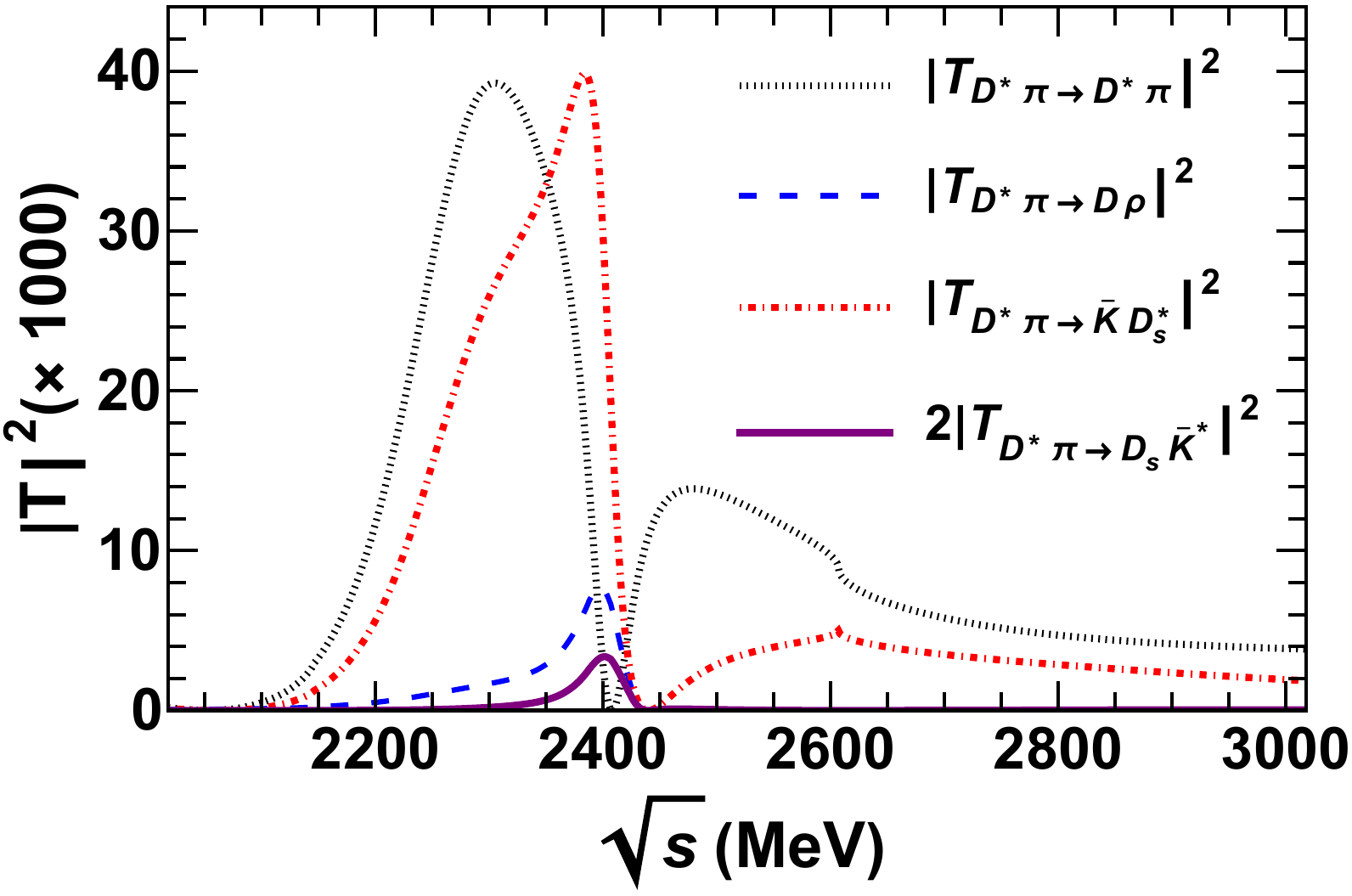}
    \includegraphics[width=0.45\textwidth]{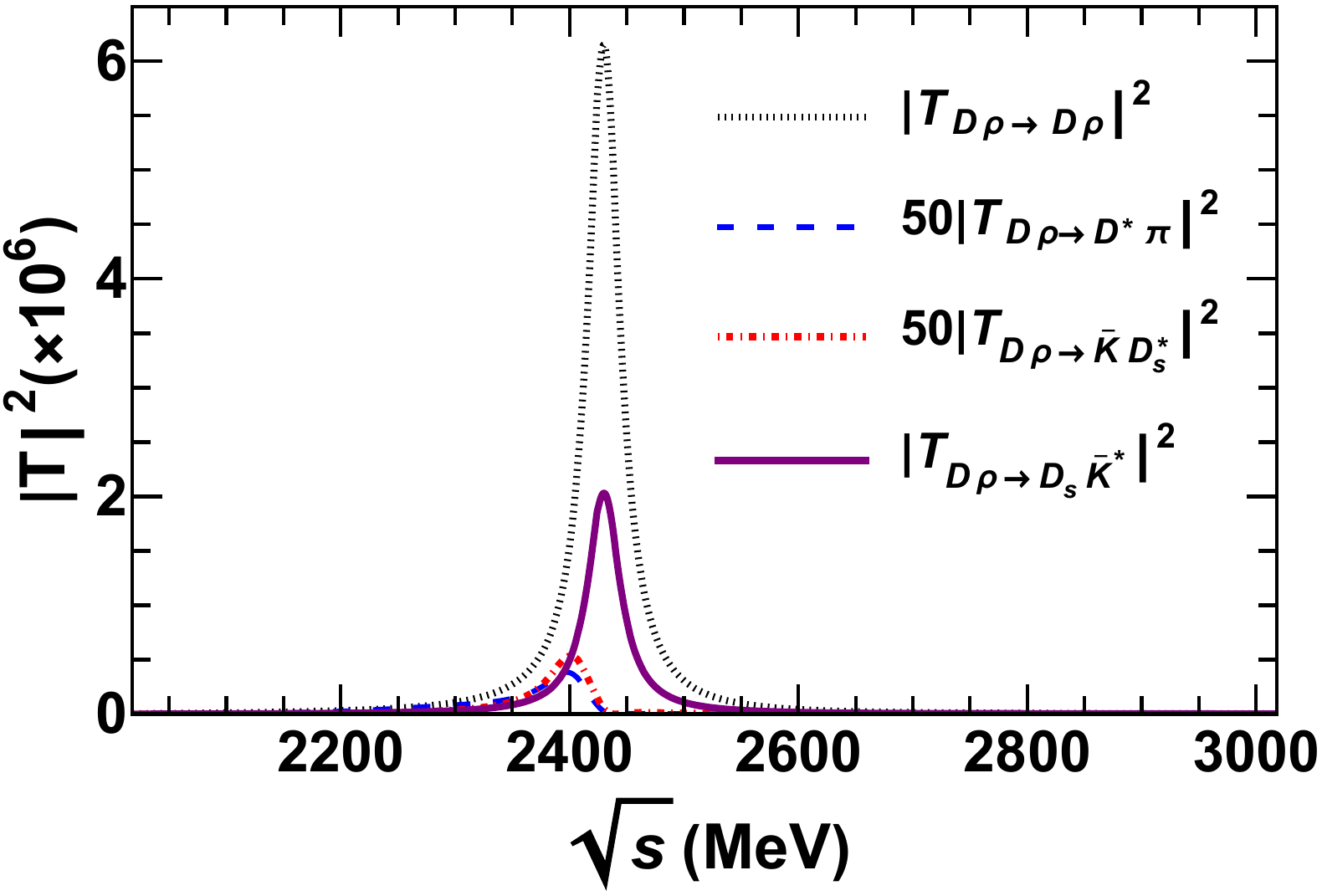}
    \\\vspace{0.5cm}
     \includegraphics[width=0.45\textwidth]{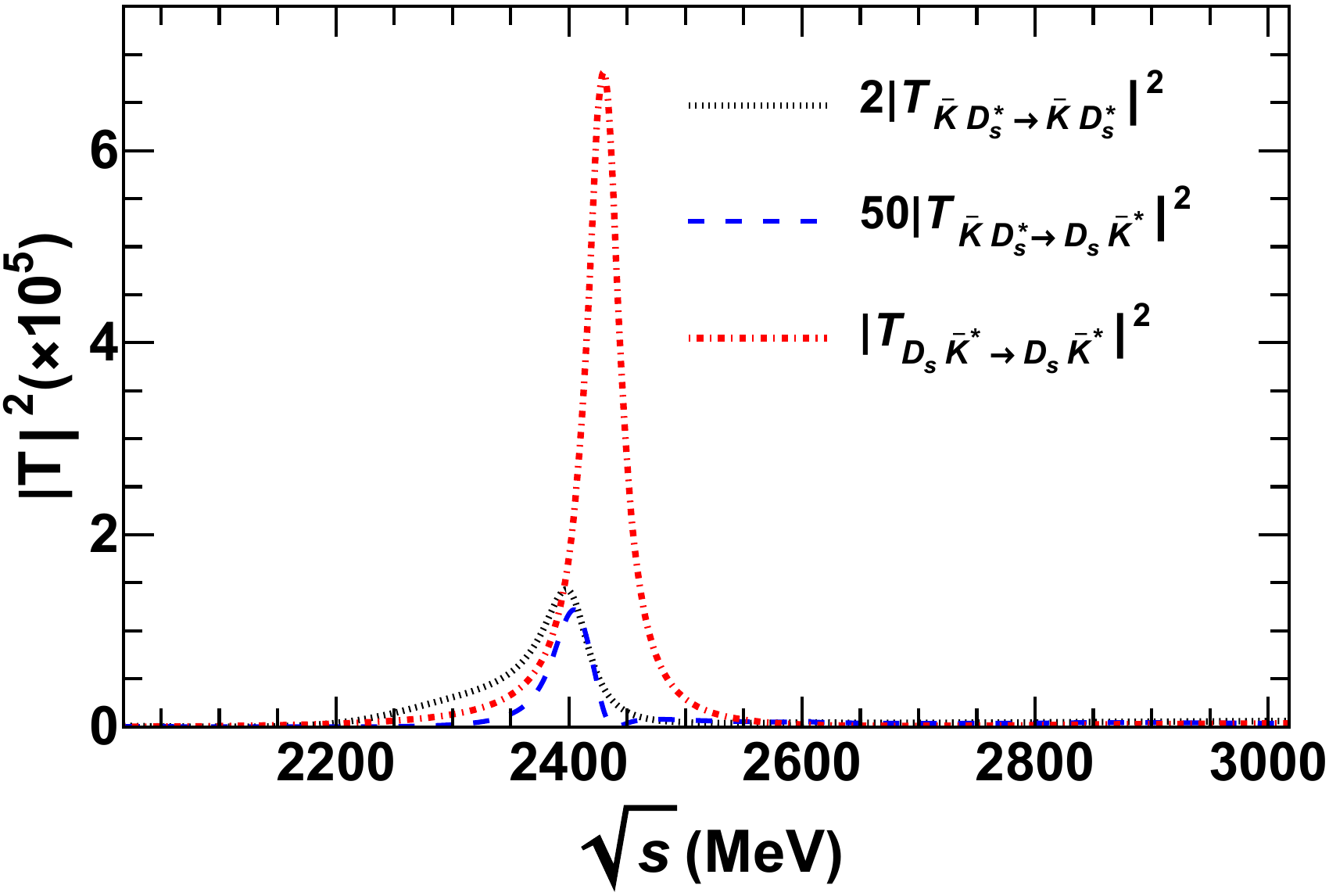}
   \caption{Squared amplitudes of different processes. A factor written in front of $|T_{...}|^2$, in the legends, implies that it has been multiplied to the squared matrix such that different amplitudes can be seen in the same scale.  }
    \label{2poleRealaxesA}
\end{figure}

The narrow pole, shown in  Fig.~\ref{twopoleOld}, remains almost unchanged and the bare quark-model pole becomes a virtual one (appearing at $2448-i0$ MeV).  The wider pole, obtained at $E-i\Gamma/2=2268-i100$,  coincides with the lowest $D_1$ state found in unitarized chiral perturbation calculations~\cite{Du:2017zvv} where the free parameters of the next-leading-order term have been fixed by using lattice QCD calculations.  The results of Refs.~\cite{Abreu:2019adi,DiPierro:2001dwf} also agree with Ref.~\cite{Du:2017zvv}.  Let us show the squared amplitudes for the different channels in Fig.~\ref{2poleRealaxesA} for further discussions.

As can be seen in the left panel of the top row of Fig.~\ref{2poleRealaxesA}, the $D^*\pi$ amplitude shows a peak on the real axis around 2305 MeV, with a full width at a half maximum of around 160 MeV. Such a width is more in agreement with the lower limit determined by the BABAR collaboration~\cite{BaBar:2006ctj}.  The same amplitude shows a zero near 2400 MeV, which is an effect of the interference between the lower energy pole arising from meson-meson dynamics and the quark-model (virtual) pole. Further, it can be seen that the $D\rho$ amplitude is almost the same in Figs.~\ref{2poleRealaxes} and \ref{2poleRealaxesA}. Thus, clearly, the $D^*\pi$ amplitude is dominated by the wider $ D_1$ state and does not show any clear sign of $D_1(2420)$, while the $D\rho$ amplitude shows only the presence of the narrow pole. In fact, we find that the $D^*\pi$ amplitude gets very little contribution from the coupled channel interactions.  Some of the transition amplitudes, like $D^*\pi \to D\rho$, do show the presence of an interference effect between a bump and a narrow peak. Such findings are in consonance with the couplings found for the two $D_1$ states to the different channels, as listed in Table~\ref{scatA}. These couplings have been determined by calculating the residues of the $t$-matrices in the complex energy plane. For the sake of completeness, we also present the position and couplings of the virtual pole in Table~\ref{scatA}.

We find it relevant to provide the values of the scattering length for different channels as well (see Table~\ref{scatA}). As discussed earlier, the value for  $D^*\pi$ is especially interesting since it can be compared with those given in Eqs.~(\ref{alat}) and (\ref{aAlice}). 
\begin{table}[ht!]
\caption{Values of the isospin 1/2 scattering lengths and the couplings, represented as $g$, of the two states $D_1$  for the different channels, as obtained in model A. Further,  we show the couplings of the virtual pole found in the work (as discussed in the text). The asterisks on $g$ indicate that the values of the pole (in the subscript) and the couplings (in the last column) correspond to a virtual pole.  }\label{scatA}
\begin{ruledtabular}
\begin{tabular}{ccccc}
& $a^{(1/2)}$& $g_{D_1(2430)}$&$g_{D_1(2420)}$& $g^{**}_{2448-i0}$ \\
&\text{(fm)}&(MeV)&\text{(MeV)}&(MeV)\\\hline
$D^*\pi$& $-0.20$ & $7250-i4995$&$-233+i5$&$302-i6826$\\
$D\rho$& $0.44-i0.18$&$-521-i355$&$15144+i356$&$-284+i719$\\
$\bar K D_s^*$&$0.00-i0.12$&$4534-i3612$&$-247-i177$&$346-i4231$\\
$D_s\bar K^*$&$0.00-i0.12$&$2-i55$&$-8739+i90$&$41-i110$\\
\end{tabular}
\end{ruledtabular}
\end{table}

At this point, we find it useful to make a brief discussion on how the couplings (for physical poles) listed in Table~\ref{scatA} can be used in determining observables.  For instance, these couplings can lead to partial decay widths, through 
\begin{align}
\Gamma_{D_1\to m_1 m_2}= \frac{p_{c.m.}}{8\pi M_{D_1}^2} \left|g_{D_1}\right|^2,\label{pwidth}
\end{align}
 where $m_1$, $m_2$ represent mesons in the decay channel, $p_{c.m.}$ refers to the center of mass momentum in the final state, $M_{D_1}$ is the mass of the $D_1$ state under consideration, and  $g_{D_1}$ is the coupling given in Table~\ref{scatA}.  Using  Eq.~(\ref{pwidth}), and considering the $\Gamma_{\text{total}}=160$ MeV for $D_1(2430)$ (as found in model A), we  obtain $\Gamma_{D_1(2430)\to D^*\pi}/\Gamma_{\text{total}} \sim 90\%$.  Note that, considering a fixed value of $M_{D_1}$,  the only open decay channel  is $D^*\pi$ and the expected branching ratio $\Gamma_{D_1(2430)\to D^*\pi}/\Gamma_{\text{total}}$ should have been 100$\%$. 
 
To obtain more precise values, we must take into account that $M_{D_1}$ varies in the range allowed by the related finite width of the decaying particle (as shown in Ref.~\cite{Oller:1997ti})  and calculate the partial decay width as 
   \begin{align}
 \Gamma_{D_1\to D^*\pi}=-\frac{1}{16 \pi^2}\int\limits_{m_{D^*}+m_{\pi}}^\infty dW \frac{p_{c.m.}}{W^2} 4 M_{D_1} \Im\left\{t_{D^*\pi \to D^*\pi}\right\}.\label{pwidth2}
 \end{align}
Such considerations are important for channels which are closed for decay at the nominal mass of a  $D_1$ state under consideration. In such cases, a finite value can be found when considering the width of the decaying state. Using  Eq.~(\ref{pwidth2}), we obtain, for instance, the branching fractions: (1) for  $D_1(2430)\to D^*\pi$ as 97.5$\%$,  (2) for  $D_1(2430)\to \bar K D_s^*$ as 2.5$\%$ and (3)  for $D_1(2430)\to D^*\rho \sim 0$.

\subsubsection{Model B}\label{modelB}
\begin{figure}[h!]
    \centering
    \includegraphics[width=0.49\textwidth]{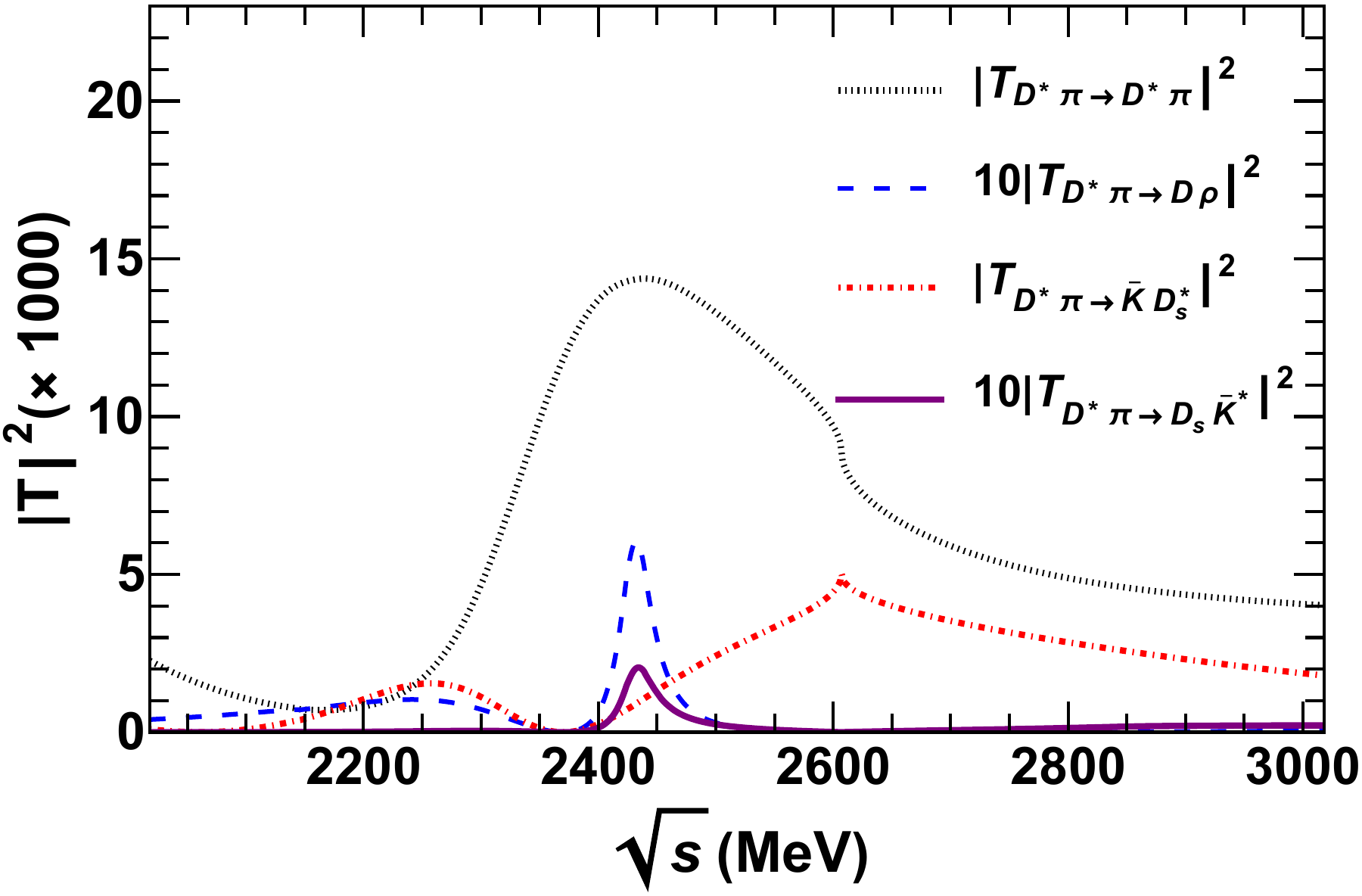}
    \includegraphics[width=0.49\textwidth]{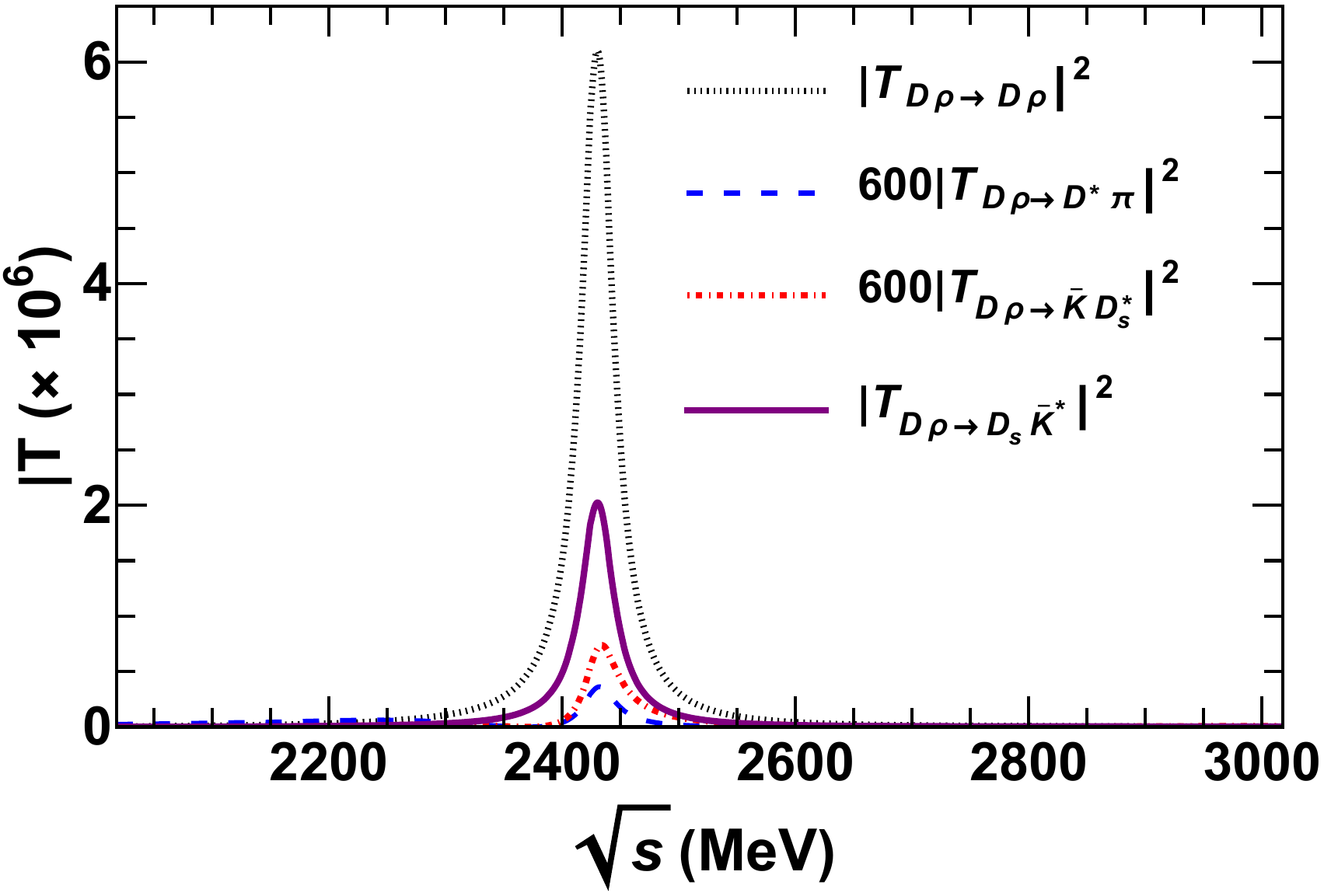}
    \\\vspace{0.5cm}
     \includegraphics[width=0.49\textwidth]{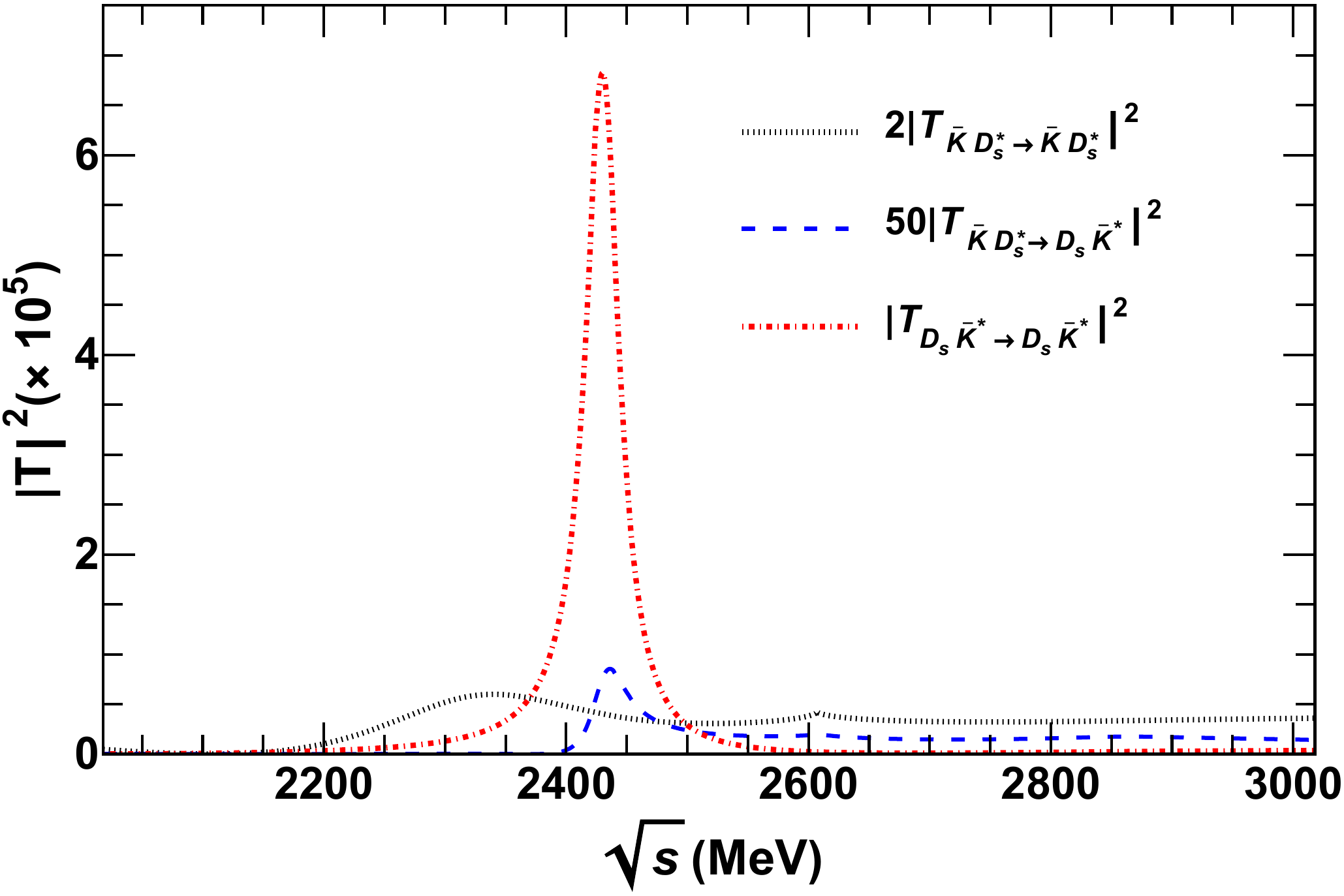}
   \caption{Same as in Fig.~\ref{2poleRealaxesA} but for the parameter choice B shown in Eq.~(\ref{vqmB}.  }
    \label{2poleRealaxesB}
\end{figure}

Contrary to Model A, where the mass of a bare pole is taken from a quark model and the coupling, as well as the sign, are chosen so as to determine a pole consistent with Refs.~\cite{Du:2017zvv,Abreu:2019adi,DiPierro:2001dwf}, we now consider the parameters of Eq.~(\ref{vqm}) as free.  We allow them to vary such as to move the lower energy pole shown in Fig.~\ref{twopoleOld} deeper in the complex plane, thus, providing the possibility of associating a bigger width to the state to be related to $D_1(2430)$.
As a result, we obtain the following parametrization
\begin{align}
V_{QM}=-\frac{10000^2}{s-2370^2}.\label{vqmB}
\end{align}
The precise position of the pole related to $D_1(2430)$  is found to be $2281-i218$ MeV. In this case, a broad bump is found on the real axis around 2436 MeV with a full width at half maximum being $\sim$ 311 MeV (see Fig.~\ref{2poleRealaxesB}). 
In this case, there occurs a constructive interference between the pole arising from the meson-meson dynamics and the bare quark model pole which becomes a virtual pole when the scattering equation is solved. The resulting mass and width values, determined from the peak on the real axis, are more in agreement with those found by the LHCb and Belle collaborations~\cite{LHCb:2015tsv,Belle:2003nsh} [as also given in Eq.~(\ref{pdgvalues})].  

It can also be noticed that the $D\rho\to D\rho$ as well as $D\rho\to D_s\bar K^*$ amplitudes are almost the same as obtained in model A (see Fig.~\ref{2poleRealaxesA}), though the strength of the transition of  $D\rho$ to the other two channels has diminished. Besides such changes, a cusp effect is seen near 2607 MeV, especially in $D^*\pi\to D^*\pi$ as well as $D^*\pi\to \bar K D_s^*$, which corresponds to the opening of the $\bar K D_s^*$ channel. We must also mention that the $D^*\pi$ amplitude, as mentioned in the discussions of model A, gets little contribution from the coupled channel interactions.

In this case, as shown in Table~\ref{scatB}, the scattering length of $D^*\pi$ turns out to be more in agreement with the value determined by the Alice collaboration [given in Eq.~(\ref{aAlice})]. We provide the couplings of the two states to the different channels, as well as to the virtual pole, too in Table~\ref{scatB}.
\begin{table}[ht!]
\caption{Values of the isospin 1/2 scattering lengths and the couplings, represented as $g$, of the two states $D_1$ and of the virtual pole (marked by asterisks) for the different channels, as obtained in model B. }\label{scatB}
\begin{ruledtabular}
\begin{tabular}{ccccc}
& $a^{(1/2)}$& $g_{D_1(2430)}$&$g_{D_1(2420)}$&$g^{**}_{2370-i0}$\\
&(fm)&(MeV)&(MeV)&(MeV)\\\hline
$D^*\pi$& $0.1$ & $5199-i3577$&$92-i105$&$-1489-i5244$\\
$D\rho$& $0.45-i0.18$&$-248-i91$&$14987+i232$&$-182+i870$\\
$\bar K D_s^*$&$0.00-i0.12$&$3806-i2249$&$186-i39$&$45-i9329$\\
$D_s\bar K^*$&$0.00-i0.12$&$-34-i36$&$-8668+i151$&$40-i207$\\
\end{tabular}
\end{ruledtabular}
\end{table}

\begin{figure}[h!]
    \centering
    \includegraphics[width=0.5\textwidth]{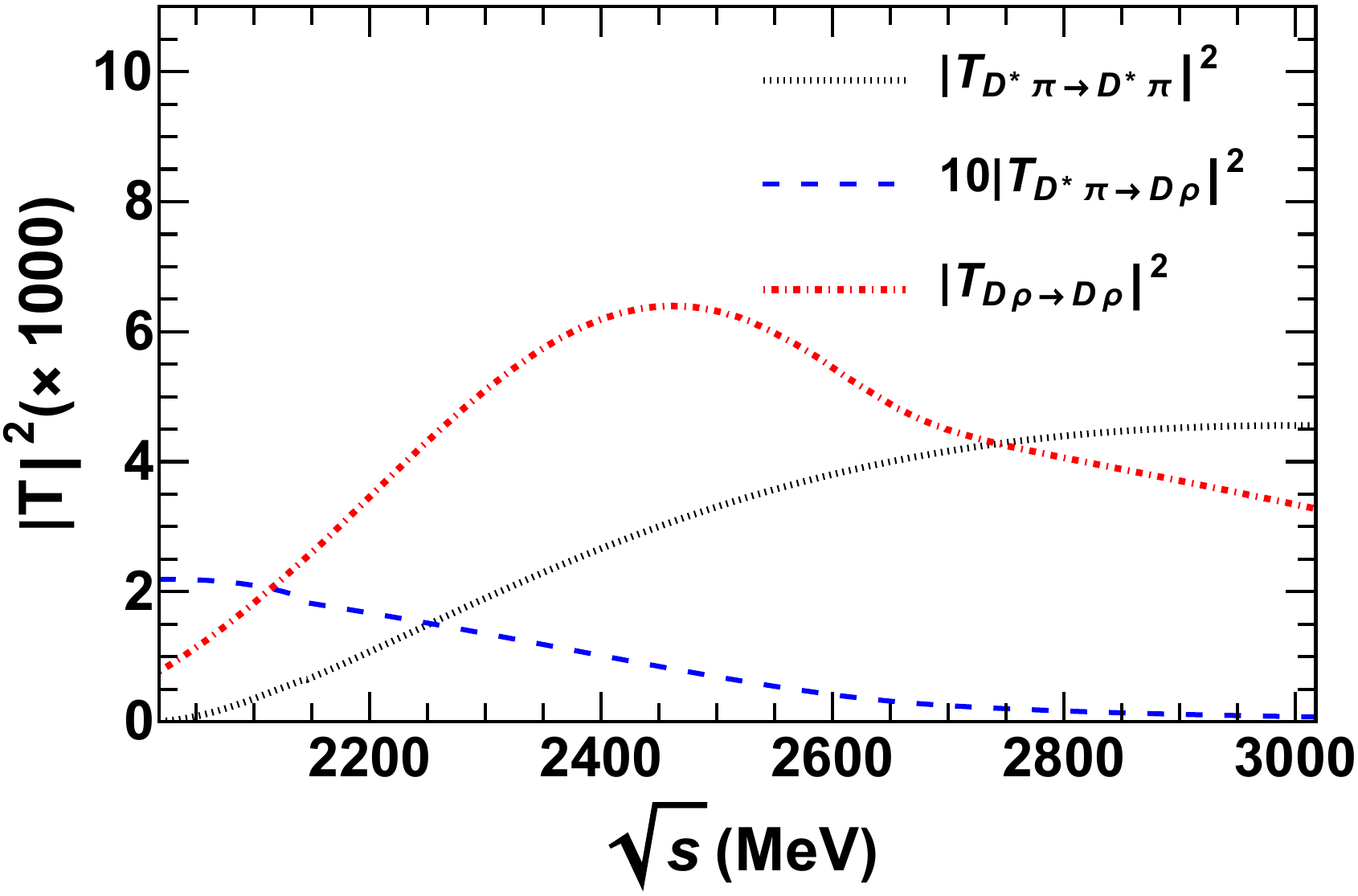}
    \caption{Squared amplitudes for the different channels in the isospin 3/2 basis. These amplitudes have been obtained by using Eq.~(\ref{tamp}) with the values of $C_{ij}$ given in the caption of Table~\ref{Table:cij}. }
    \label{amp3h}
\end{figure}

To summarize this section, we can say that we have studied the interactions of different meson-meson systems coupling to the quantum numbers of $D_1$ states. We find that, within the model considered, the meson-meson interactions can well describe the properties of $D_1(2420)$. A wider pole at lower energies is also generated from the interactions, though the mass and width are not found to be in good agreement with the known properties of $D_1(2430)$. We find that adding a bare quark model pole to the $D^* \pi$ amplitude improves the situation. We present two scenarios, which lead to values of $D^*\pi$ scattering length in agreement with the conflicting ones known for $D\pi$ from lattice QCD-inspired models and from the Alice collaboration. We now study how such scenarios reflect in terms of the correlation functions. To calculate correlation functions in particle basis, we require the amplitudes in the isospin 3/2 basis too. We end this section by showing such amplitudes in Fig.~\ref{amp3h}.
 Notice that the interactions are weakly repulsive in this isospin configuration and, thus, no states are formed in this case. The scattering lengths in this case are $a_{D^*\pi}^{(3/2)} =0.1$ fm and 
$a_{D\rho}^{(3/2)} =0.2$ fm. The value of the isospin 3/2 scattering length, for the $D^*\pi$ channel, is in agreement with the ones summarized in Eqs.~(\ref{alat}) and (\ref{aAlice}). We remind the reader that our sign convention (as given in Sec.~\ref{scatlen}) is opposite to the one followed in the works leading to the values of Eqs.~(\ref{alat}) and (\ref{aAlice}).

\section{Correlation Functions}\label{sec-corrfunction}
As stated earlier the purpose of our work is to determine the correlation function for the $D^*\pi$ system. We focus on investigating the $D^{*+(0)}\pi^{0(+)}$ system where Coulomb interactions are absent and strong interactions dominate. The idea is also to study the dependence of the correlation function on the size of the source, such as to find if experimental data on processes other than that studied in Ref.~\cite{ALICE:2024bhk} can bring useful information on the topic.

\subsection{Formalism}\label{subsec-corrfunction-formalism}

The femtoscopic analysis is based on the estimation of the correlation functions (CFs). A two-particle correlation function is constructed as the ratio of the probability of measuring the two-particle state and the product of the probabilities of measuring each individual particle~\cite{Lisa:2005dd}.  A convenient form relating the correlation function to the source function by means of a convolution with the relative two-particle wave function $\Psi$ is written, after certain approximations, as~\cite{Lisa:2005dd,Koonin:1977fh,Pratt:1986cc,Lednicky:1981su,Lednicky:1998}
\begin{eqnarray}
C(k) & = & \int d^3 r S_{12}(\vec{r}) \vert \Psi (\vec{k} ; \vec{r}) \vert ^2, 
\label{cf1}
\end{eqnarray}
where $ \vec{k} $ is the relative momentum in the c.m. of the pair; $\vec{r}$ is the relative distance between the two particles; and $ S_{12}(\vec{r}) $ is the normalized source function, $\int d^3 r S_{12}(\vec{r}) = 1$, describing the distribution of relative positions of particles with identical velocities as they move in their asymptotic state (for a detailed discussion see for example Ref.~\cite{Lisa:2005dd}). As a consequence, the expression above for $C(k)$ encodes information on both the hadron source and the hadron-hadron interactions and is commonly named as Koonin–Pratt equation~\cite{Koonin:1977fh,Pratt:1986cc}~\footnote{Eq.~(\ref{cf1}) is also called by some authors like those from Refs.~\cite{Kamiya:2021hdb} as Koonin–Pratt–Lednicky–Lyuboshits–Lyuboshits formula due to subsequent contributions~\cite{Lednicky:1981su,Lednicky:1998}}. 

In the present work, we employ a source function parametrized as a static Gaussian normalized to unity, i.e.,
\begin{eqnarray}
S_{12}(\vec{r})  & = & \frac{1}{\left(  4 \pi \right)^{\frac{3}{2}} R^3} \exp{\left(  -\frac{r^2}{4 R^2 }\right)} ,  
\label{sourcef1}
\end{eqnarray}
where $R$ is the source size parameter. As discussed in Ref.~\cite{Lisa:2005dd},  Gaussian parametrizations provide an acceptable minimal description of data in a much more simpler way than others with non-Gaussian aspects of the correlation, such as the ones based on the decomposition in spherical or Cartesian harmonics. Thus, the source function in Eq.~(\ref{sourcef1}) can be seen as the appropriate parametrization for the sake of its functionality. 

To connect the CF to the coupled-channel approach described in the previous section, we adopt the framework summarized in Refs.~\cite{Vidana:2023olz,Feijoo:2023sfe,Albaladejo:2023pzq,Liu:2022nec,Liu:2023uly,Liu:2023wfo}, in which the generalized coupled-channel CF for a specific channel $i$ reads
\begin{eqnarray}
C_i(k) & = & 1 + 4 \pi \theta (q_{max} - k) \int_{0}^{\infty} d r r^2 S_{12}(\vec{r}) \left( \sum_j w_j \vert j_0(kr) \delta_{ji} + T_{ji}(\sqrt{s}) \widetilde{G}_j(r; s) \vert^2 - j_0^2(kr) \right) ,  \nonumber \\
\label{cf2}
\end{eqnarray}
where $ w_j $ is the weight of the observed channel $ j $ (we use $ w_j = 1$); $ j_{\nu}(kr) $ is the spherical Bessel function; $E = \sqrt{s}$ is the c.m. energy; the relative momentum of the channel is $k = \lambda^{1/2} (s,m_1^2,m_2^2)/(2\sqrt{s})$ ($\lambda$ being the K\"allen function and $m_1, m_2$ the masses of the mesons in the channel $i$); $ T_{ji} $ are the elements of the scattering matrix encoding the meson–meson interactions, obtained and analyzed in the previous section; and the $\widetilde{G}_j(r; s)$ function is defined as 
\begin{eqnarray}
\widetilde{G}_j(r; s) & = & \int_{\vert \vec{q} \vert < q_{max} } \frac{d^3 q}{(2\pi)^3} \frac{\omega_1^{(j)} + \omega_2^{(j)} }{2 \omega_1^{(j)} \omega_2^{(j)} } \frac{j_0(qr)}{s - \left( \omega_1^{(j)} + \omega_2^{(j)} \right)^2 +i \varepsilon} ,   
\label{gtilde}
\end{eqnarray}
with $ \omega_a^{(j)} \equiv \omega_a^{(j)}(k) = \sqrt{k^2 + m_a^2}$ being the energy of the particle $a$, and $q_{max}$ being a sharp cutoff momentum introduced to regularize the $r \to 0$ behavior.  We choose  a value for $q_{max}$ within its natural range ($ [600,900] \ \mathrm{MeV}$): $q_{max} = 700 \  \mathrm{MeV}$. We remark that the results for the CFs remain almost the same for different values of $q_{max}$ within the mentioned range, as expected because of the presence of $j_0(qr)$ in the integrand, which prevents sizable changes for large values of $q$. 

\subsection{Lednicky-Lyuboshits approximation }\label{CLLaprox}

To shed some light on the interpretation of the CFs, it can be instructive to review the Lednicky-Lyuboshits (LL) model, which is based on replacing the full wave function for a single channel by its nonrelativistic, asymptotic ($r\to \infty $) form, corresponding to the superposition of plane and converging spherical waves~\cite{Lednicky:1981su}. In particular, we benefit from the discussion presented in the Appendix of Ref.~\cite{Albaladejo:2023pzq} and Secs. V.B and V.C of Ref.~\cite{Kamiya:2021hdb}, which have some of their fundamental aspects reproduced here.

Proceeding ahead, the consideration of the LL approximation, together with a Gaussian source, and using the relationship between the standard quantum mechanics amplitude $ f(k) $ and the scattering matrix $ T $, i.e., $ f(k) = -T/(8 \pi \sqrt{s}) $, allow us to write the single-channel CF as~\cite{Fabbietti:2020bfg,Lednicky:1981su,Kamiya:2021hdb,Albaladejo:2023pzq}
\begin{eqnarray}
C_{LL}(k) & = & 1 + \frac{ \vert T \vert ^2 }{2 R^2 (8 \pi \sqrt{s})^2} F_1\left( \frac{r_{eff}}{R} \right) -  \frac{2 Re[T] }{ 8 \pi^{3/2} R \sqrt{s}} F_2(2 k R) + \frac{ Im[T] }{ R \sqrt{s}}F_3(2 k R) ,  
\label{cfLL0}
\end{eqnarray}
where $F_1(z) = 1 - z/(2\sqrt{\pi})$, $F_2(z) = \int_0^z dt e^{t^2 - z^2}/z$ and $F_3(z) = ( 1 - e^{ - z^2})/z$; $r_{eff}$ is the effective range.
An alternative version of Eq.~(\ref{cfLL0}) can be obtained by employing the formula~\footnote{Once again, we emphasize that our sign convention is different from that of Ref.~\cite{Albaladejo:2023pzq}, i.e. $\lim_{k \to 0} k \cot{\delta(k)} \equiv - a ^{-1}$, which gives a different sign in the last term between parentheses.} 
\begin{eqnarray}
- \frac{T}{8 \pi \sqrt{s} } \equiv \frac{1}{k \cot{\delta(k)} - i k}  =  \frac{R}{ - R/a - i k R} ,
\label{TLL0}
\end{eqnarray}
where $r_{eff}$ is taken as zero.

In this way Eq.~(\ref{cfLL0}) becomes~\cite{Kamiya:2021hdb,Albaladejo:2023pzq}
\begin{eqnarray}
C_{LL}(x,y) & = & 1 + \frac{1}{x^2+y^2} \left[ \frac{1}{2} - \frac{2 y}{\sqrt{\pi}} F_2(2 x) - x F_3 (2x) \right] ,  
\label{cfLL1}
\end{eqnarray}
where $x= kR$ and $y=R/a$. In the low-momentum limit $(x \to 0)$ we have $F_2  \to 1 $ and  $F_3 \to 0 $, which yields 
\begin{eqnarray}
C_{LL}(x,y) & \overset{x \rightarrow 0}{\longrightarrow} & 1 - \frac{2}{\pi} + \frac{1}{2} \left( \frac{1}{y} - \frac{2 }{\sqrt{\pi}} \right)^2  \nonumber \\
& = & 1 - \frac{2}{\sqrt{\pi}} \left( \frac{a}{R} \right) + \frac{1}{2} \left( \frac{a}{R} \right)^2.  
\label{cfLL2}
\end{eqnarray}

\begin{figure}[ht!]
    \centering
\includegraphics[width=0.45\textwidth]{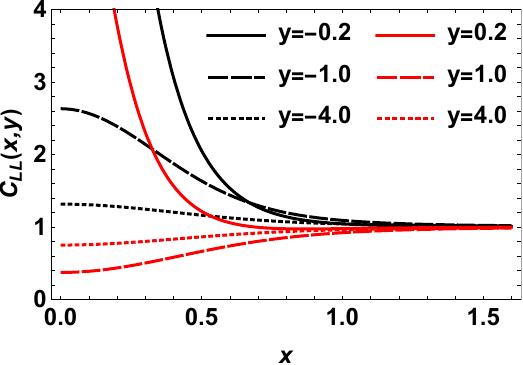}
\includegraphics[width=0.45\textwidth]{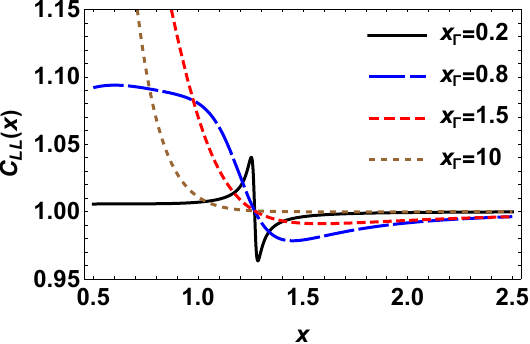}
    \caption{Left panel: CF in LL approximation given in Eq.~(\ref{cfLL1}) as a function of $x= kR$, taking different values of $y=R/a$. Right panel:  CF in LL approximation given in Eq.~(\ref{cfLL3}) as a function of $x= kR$, taking different values of $x_{\Gamma} = \sqrt{\mu \Gamma R^2} $. We have considered here the reduced mass $\mu$ of the $D^* \pi$ system, and the value of $x_R = k_R R \simeq 1.3$, which is associated to a resonance localized at $k_R \simeq 250 $ MeV and $R = 1$ fm. The value of $x_{\Gamma} \simeq 0.8 $ corresponds to a resonance with $\Gamma \simeq 200 $ MeV. }
    \label{fig:cfll}
\end{figure}

Thus, considering an attractive interaction generated by the strong force, the CF given in Eqs.~(\ref{cfLL1}) and~(\ref{cfLL2}) behaves as follows. Near the threshold, for negative values of the scattering length (which means an unbound scenario for the system) the CF acquires (i) a strong enhancement  when $|a| \gg R $ (i.e., smaller source), (ii) a moderate enhancement when $|a| \sim R $, and (iii)  a value $C_{LL}(0) \gtrsim 1 $  when $|a| \ll R $ (larger source). On the other hand, in the situation of $a>0$, corresponding to an attractive interaction which could generate a bound or quasibound state, in the low-momentum limit, the CF achieves (i) a strong enhancement for smaller sources $(a \gg R )$, (ii) a value very close to one when $a/R \sim 4/ \sqrt{\pi } \sim 2.3 $, (iii) its minimum value  ($\simeq 0.4$) when $ a / R \sim 2 / \sqrt{\pi} \simeq 1.1 $,  (iv) a moderate dip when $a < R $; and (v) a value $  C_{LL}(0) \lesssim 1 $  for larger sources ($a \ll R $).  
This behavior is summarized in Fig.~\ref{fig:cfll} (left panel), where one can see that the enhancement of CF at a given value of $R$ is not conclusive concerning the formation of a bound or quasibound state. Notwithstanding this,  as a consequence of the dependence of the CF on $R$ and $a$, one can infer the existence of a bound or quasibound state when, near the threshold, the CF moves from an enhancement to a dip at $a \simeq R$ as $R$ increases. In this sense, experimental analyses of the CF in systems with different sizes, for instance, $pp, pA $, and $AA$ collisions, deserve special attention.

Pursuing further the analysis, to understand the effect of the presence of a resonance at a given momentum $ k_R $ in the CF, it is more convenient to use Eq.~(\ref{TLL0}) and write Eq.~(\ref{cfLL0}) in the form
\begin{eqnarray}
C_{LL}(x) & = & 1 + \frac{ \sin^2{\delta(k)}}{2 x^2} \left( e^{-4 x^2} + \frac{4 x F_2(2 x)}{\sqrt{\pi}} \cot{\delta(k)} \right) ,  
\label{cfLL3}
\end{eqnarray}
Then, considering that a resonance present at $k_R$ generates $ \delta (k_R)  = \pi / 2$, which when used in Eq.~(\ref{cfLL3}) yields $C_{LL}(x_R) = 1 + \frac{ e^{-4 x_R^2}}{2 x_R^2}; \ x_R = k_R R $. Accordingly, the CF ends up having the following properties (i) $C_{LL}^{\prime}(x_R) < 0 $, (ii)  $C_{LL} < 1 $ for $x \gtrsim x_R$, and (iii)  $C_{LL} \simeq 1$ for large $x_R$. To be more didactic, from the analytical expression for the Breit-Wigner-like phase shift, $\delta (E) = \Gamma (E)/(2E_R-2E)$ ($\Gamma$ being the width and $E_R$ the energy at $k_R $), one gets $\cot{\delta (k)} \sim - (x^2 - x_R^2)/x_{\Gamma}^2$, with $x_{\Gamma}^2 = \mu \Gamma R^2 $ ($\mu$ being the reduced mass of the two particles in the channel). Thus, the use of this last expression of $\cot{\delta (k)}$ in Eq.~(\ref{cfLL3}) can engender in $C_{LL}$ (i) a maximum at $k \lesssim k_R$ and a pronounced minimum at $k \gtrsim k_R$ for small $x_{\Gamma}$, (ii) a weakened minimum at $k \gtrsim k_R$ for intermediate values of $x_{\Gamma}$, and (iii) an almost plateaulike appearance at $k \lesssim k_R$ for large $x_{\Gamma}$. This behavior is summarized in Fig.~\ref{fig:cfll} (right panel) in accordance with that in Ref.~\cite{Albaladejo:2023pzq}.

In the end, the form and what can be interpreted from the CF are strongly dependent on the parameters, namely $k_R, \Gamma, R$, and $a$. We also remark that the application of the LL approximation in the interpretation of the present context must be seen with caution since we treat a coupled-channel problem, which is naturally more complex to analyze and understand. In this sense, it will be helpful to perform comparisons between the single-channel LL and coupled-channel CFs in order to get insights into the reliability of the LL model in the description of this problem.

\subsection{Results}\label{subsec-corrfunction-results}

\subsubsection{CFs in isospin basis}

\begin{figure}[ht!]
    \centering
\includegraphics[width=0.32\textwidth]{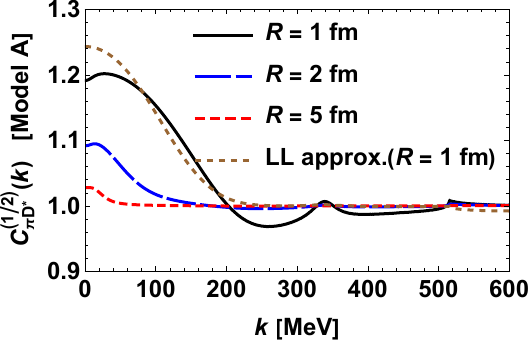}
\includegraphics[width=0.32\textwidth]{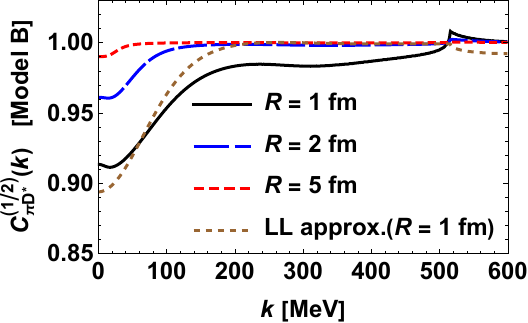}
\includegraphics[width=0.32\textwidth]{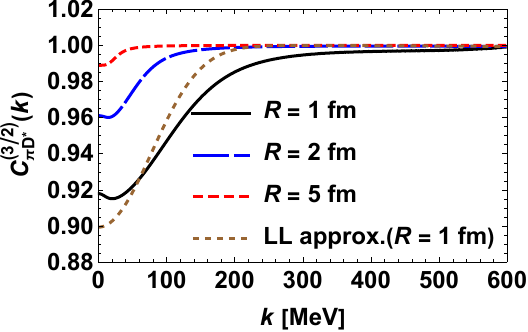} \\
\includegraphics[width=0.32\textwidth]{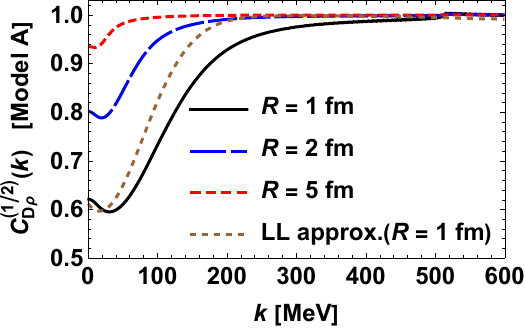}
\includegraphics[width=0.32\textwidth]{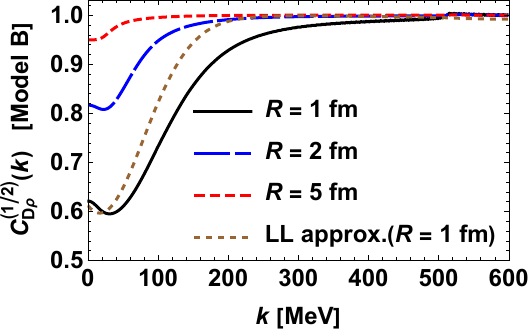}
\includegraphics[width=0.32\textwidth]{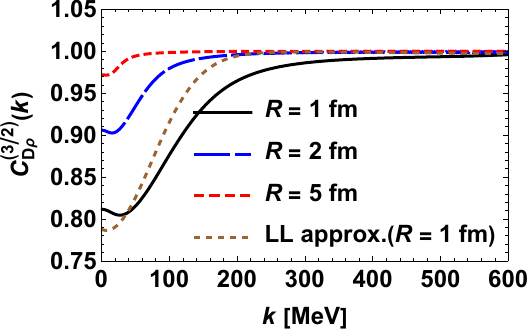} 
    \caption{CFs for the $D^* \pi $ (top panels) and $D \rho $ (bottom panels) channels as functions of their relative momentum $k$, taking different values of the source size. The left, center, and right panels show the results obtained, respectively, with $I=1/2$ considering the model A, with $I=1/2$ considering model B, and with $I=3/2$. Recall that the interaction in the isospin 3/2 configuration is repulsive and weaker as compared to the case of isospin 1/2. Thus, no states are formed in the isospin 3/2 case and the amplitudes are obtained from the same model [using Eq.~(\ref{tamp})].  }
    \label{fig:cf12}
\end{figure}

In Fig.~\ref{fig:cf12} we plot the results of the correlation functions for the most relevant channels $D^* \pi $ and $D \rho $ as functions of their c.m. momentum $k$, for different values of the range parameter $R$ of the source, considering the following scenarios: $I=1/2$ with model A,  $I=1/2$ with model B and $I=3/2$. For the sake of comparison and to reach a more profound comprehension concerning our findings, the results with the single-channel LL approximation are also included.

\begin{figure}[ht!]
    \centering
\includegraphics[width=0.32\textwidth]{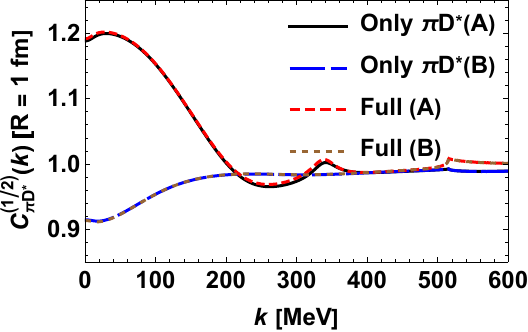}
\includegraphics[width=0.32\textwidth]{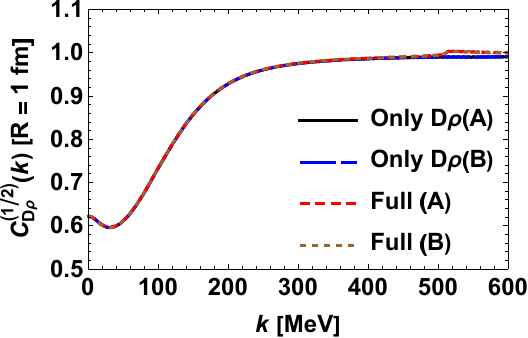}
    \caption{CFs with $I=1/2$ for the $D^* \pi $ (left panel) and $D \rho $ (right panel) as functions of their relative momentum $k$ considering the single contribution of the elastic channel [i.e., only $i=j$ of the sum in Eq.~(\ref{cf1})], and the total CF.  }
    \label{fig:cf12singlefull}
\end{figure}

First, one can notice the distinct behavior of the $C_{D^* \pi}^{(1/2)}(k)$ when models A or B are employed ( see left and center panels in the upper row of Fig.~\ref{fig:cf12}). To start with the discussions, let us focus on the curves related to the smallest source size parameter, $R=1 \ \mathrm{fm}$. In the case of model A, at threshold, we have $C_{D^* \pi}^{(1/2)}(k=0) > 1$, because of the attractive character of this channel and the negative scattering length. In the sequence, as $k$ increases a moderate minimum and a bump are found in the region $220 \lesssim k \lesssim 360 \ \mathrm{MeV}$. Interestingly, these effects reflect essentially the behavior of the $T_{D^* \pi,D^* \pi}^{(1/2)}$ amplitude, since the other contributions $T_{D^* \pi,D \rho}^{(1/2)}$ and $T_{D^* \pi, \bar K D_s^*}$ are negligible, as shown in Fig.~\ref{fig:cf12singlefull}~\footnote{This finding is in agreement with the result on the $D^*\pi$ amplitude discussed in section~\ref{modelA} and \ref{modelB}.  An equivalent effect is also found in the femtoscopic analysis of the coupled-channel $N \Xi$ and $\Lambda\Lambda$ interactions~\cite{Kamiya:2021hdb}.}. In this sense,   the minimum (bump) at $k \gtrsim 250 \ \mathrm{MeV}$ ($k \simeq 340 \ \mathrm{MeV}$) is associated to the broad peak (dip) in $T_{D^* \pi,D^* \pi}^{(1/2)}$ at $\sqrt{s} \sim 2304 \ \mathrm{MeV}$ ($\sqrt{s} \sim 2405 \ \mathrm{MeV}$). Thus, according to our model, the CF encodes the manifestation of the interference between the poles present in $T_{D^* \pi,D^* \pi}^{(1/2)}$ (see discussions in Sec.~\ref{modelA}). However, these effects are no longer prominent for larger values of the source size parameter. Also, a cusp at $k \simeq 518 \ \mathrm{MeV}$ is seen and comes from the effect of the $\bar K D_s^*$ threshold. 
Notably, when compared to the single-channel LL results, these CFs have a similar qualitative behavior only near the threshold,  having values larger than one. The $ C_{LL} $ does not acquire a minimum from the peak in $T_{D^* \pi,D^* \pi}^{(1/2)}$, possibly because of its large width (as argued in Sec.~\ref{CLLaprox}).  

On the other hand, at the threshold model B generates $C_{D^* \pi}^{(1/2)}(k=0) \lesssim 1$, which is compatible with the result expected within the LL approximation when $a_{D^* \pi}^{(1/2)} = 0.1 \ \mathrm{fm} < 2.3 R$. After that, the CF slightly increases with $k$, and presents almost a plateau shape, which comes from the interference between the states discussed in Sec.~\ref{modelB}; then it shows also a cusp at $k \simeq 518 \ \mathrm{MeV}$  and goes to one. As in the former model, in this case too the full CF expresses the behavior of the $T_{D^* \pi,D^* \pi}^{(1/2)}$ amplitude (see the left panel of Fig.~\ref{fig:cf12singlefull}); and is qualitatively similar to the single-channel LL results only near the threshold, since the $ C_{LL} $ goes faster  towards unity.

For the $I=3/2$ channel (top, right panel of Fig.~\ref{fig:cf12}),  at threshold $C_{D^* \pi}^{(3/2)}(k)$ starts moderately lower than one. This, considering the value of the scattering length $a_{D^* \pi}^{(3/2)} = 0.1$ fm, is compatible with the behavior expected within the LL approximation when  $a_{D^* \pi}^{(3/2)} = 0.1 \ \mathrm{fm} < 2.3 R$.  After that, the CF increases with the augmentation of $k$ and goes to one; no other effect appears as no state is present.

Now we move on to the channel $ D \rho $ (bottom panels of Fig.~\ref{fig:cf12}), whose scattering length has an imaginary component. We do not see sizable differences in the $C_{D \rho}^{(1/2)}(k)$ obtained considering the models A and B (as expected from the similarity of the $D\rho$ amplitude in the two models).
Noticing that $Re[a_{D \rho}^{(1/2)}] = 0.44 \ \mathrm{fm} < 2.3 R$, then one can expect that  $C_{D^* \pi}^{(1/2)}(k=0) < 1 $. However,  when compared with the results for $D^* \pi$ in model B, the CF experiences a substantial dip. Taking advantage of the analysis in the previous section, this may be interpreted as the influence of the narrow state present in the $T_{D \rho,D \rho}^{(1/2)}$ amplitude below the $D \rho$ threshold, which as  shown in Fig.~\ref{fig:cf12singlefull}, provides the relevant contribution.
 When compared to the single-channel LL approximation, at threshold $C_{D \rho}^{(1/2)}(k)$ is quite near $ C_{LL} $ but goes more slowly towards unity.

\subsubsection{CFs in physical basis}

We remark that the CFs presented so far, for the relevant channels $D^* \pi $ and $D \rho $, are in the isospin basis. Therefore, in order to provide measurable CFs, we need to express them on the particle basis. In the case of $D^* \pi $ (the case of $D \rho $ is completely analogous), we consider the isospin doublet of the vector charmed meson and isospin triplet of the pion as $D \equiv (\vert D^{*+} \rangle, - \vert D^{*0} \rangle)$ and $\pi \equiv (- \vert \pi^+ \rangle, \vert \pi^0 \rangle, \vert \pi^- \rangle)$, respectively. Then, for $D^* \pi $ states with $I_3 = +1/2$, the  particle basis is given by ${ \vert D^{*0} \pi^+ \rangle , \vert D^{*+} \pi^0 \rangle } $, which is related to the isospin basis through (denoting states as $ \vert D \pi, I \rangle  $)
\begin{eqnarray}
\vert D^{*0} \pi^+ \rangle  & = & - \left[ \sqrt{\frac{2}{3}} \left\vert D \pi, \frac{1}{2} \right\rangle - \sqrt{\frac{1}{3}} \left\vert D \pi, \frac{3}{2}  \right\rangle \right], \nonumber \\
\vert D^{*+} \pi^0 \rangle  & = &  \sqrt{\frac{1}{3}} \left\vert D \pi, \frac{1}{2} \right\rangle + \sqrt{\frac{2}{3}} \left\vert D \pi, \frac{3}{2}  \right\rangle . 
\label{rel1}
\end{eqnarray}
With these last expressions, we can write the two-particle wave function  for charged states as 
\begin{eqnarray}
\Psi_{D^{*0} \pi^+ \to D^{*0} \pi^+}  & \equiv  &  \frac{2}{3} \Psi_{D^{*} \pi}^{\left(\frac{1}{2}\right)} + \frac{1}{3} \Psi_{D^{*} \pi}^{\left(\frac{3}{2}\right)}, \nonumber \\
\Psi_{D^{*+} \pi^0 \to D^{*0} \pi^+}  & \equiv  &  - \frac{\sqrt{2}}{3} \Psi_{D^{*} \pi}^{\left(\frac{1}{2}\right)} + \frac{\sqrt{2}}{3} \Psi_{D^{*} \pi}^{\left(\frac{3}{2}\right)}, \nonumber \\
\Psi_{D^{*0} \pi^+ \to D^{*+} \pi^0}  & \equiv  &  - \frac{\sqrt{2}}{3} \Psi_{D^{*} \pi}^{\left(\frac{1}{2}\right)} + \frac{\sqrt{2}}{3} \Psi_{D^{*} \pi}^{\left(\frac{3}{2}\right)}, \nonumber \\
\Psi_{D^{*+} \pi^0 \to D^{*+} \pi^0}  & \equiv  &  \frac{1}{3} \Psi_{D^{*} \pi}^{\left(\frac{1}{2}\right)} + \frac{2}{3} \Psi_{D^{*} \pi}^{\left(\frac{3}{2}\right)}, 
\label{wf1}
\end{eqnarray}
where the superscript on $ \Psi$ indicates the related isospin. As a consequence, using Eq.~(\ref{wf1}) in ~(\ref{cf1}) we get
\begin{eqnarray}
C_{ D^{*0} \pi^+}  & \equiv  & C_{D^{*0} \pi^+ \to D^{*0} \pi^+} + C_{D^{*+} \pi^0 \to D^{*0} \pi^+}   = \frac{2}{3}  C_{D^{*} \pi}^{\left(\frac{1}{2}\right)} + \frac{1}{3} C_{D^{*} \pi}^{\left(\frac{3}{2}\right)}, \nonumber \\
C_{ D^{*+} \pi^0}  & \equiv  & C_{D^{*0} \pi^+ \to D^{*+} \pi^0} + C_{D^{*+} \pi^0 \to D^{*+} \pi^0} =  \frac{1}{3}  C_{D^{*} \pi}^{\left(\frac{1}{2}\right)} + \frac{2}{3} C_{D^{*} \pi}^{\left(\frac{3}{2}\right)} .
\label{rel2}
\end{eqnarray}

\begin{figure}[ht!]
    \centering
\includegraphics[width=0.32\textwidth]{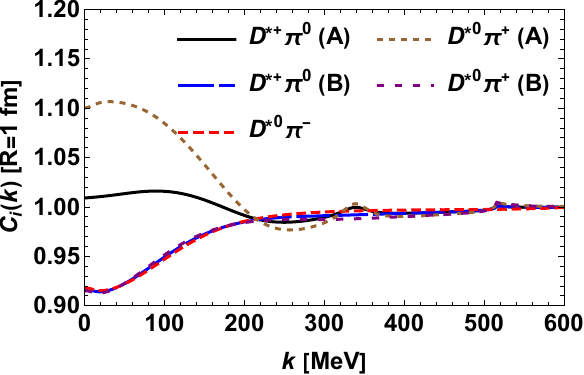}
\includegraphics[width=0.32\textwidth]{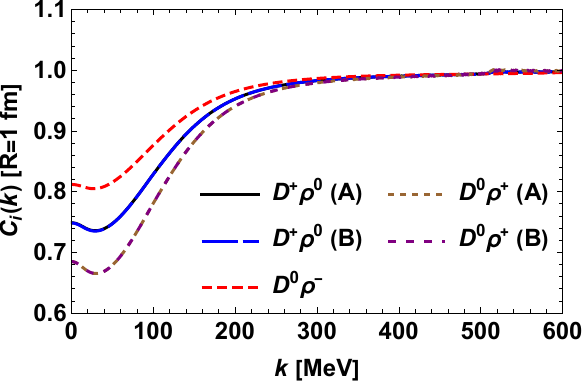}
    \caption{CFs for the physical $D^{*} \pi$ and $ D \rho $ states defined in Eq.~(\ref{rel1}) in both cases of models A and B, taking the source size parameter $R=1$ fm. The CFs $C_{ D^{*0} \pi^-}(k)$ and $C_{ D^{0} \rho^-}(k)$, equivalent to the corresponding $C_i^{(3/2)}(k)$,  are also plotted. 
}
    \label{fig:cf12partbasis}
\end{figure}

We show the CFs for the  $D^{*} \pi$ and $ D \rho $ states {in the particle basis defined in Eq.~(\ref{rel2}), in Fig.~\ref{fig:cf12partbasis}, for both models A and B. The CFs $C_{ D^{*0} \pi^-}(k)$ and $C_{ D^{0} \rho^-}(k)$ are also plotted; since these states have $I_3 = -3/2$, their corresponding CFs are naturally equal to $C_i^{(3/2)}(k)$. It can be seen that for model A the features of the  $T_{D^* \pi,D^* \pi}^{(1/2)}$ amplitude are more notable in the channel $D^{*0} \pi^+$, because of the bigger weight of the $I=1/2$ channel in its wave function. In contrast, for model B there is no sizable difference among the channels $D^{*0} \pi^+, D^{*+} \pi^0$, and $D^{*0} \pi^-$, due to the similarity among $C_{D^* \pi}^{(1/2)}(k)$ and $C_{D^* \pi}^{(3/2)}(k)$. 
Going to the scenario of $ D \rho $, the difference coming from the isospin weights produces $C_{ D^{+} \rho^0}(k)$ closer to one at threshold than $C_{ D^{0} \rho^+}(k) $.
Hence, one can conclude that the $D^{*0} \pi^+$ and $ D^{0} \rho^+ $ channels are more appropriate to test both models. It is worth mentioning that correlation functions for $D^{*+}\pi^+$ and  $D^{*+}\pi^-$ have been calculated in Ref.~\cite{Torres-Rincon:2023qll}, although with very different poles present in the amplitudes, those in line with Refs.~\cite{Abreu:2019adi,Du:2017zvv}.   

In the end, we stress the main conclusion of this study, namely: our findings suggest that $C_{D^* \pi}(k)$ and $C_{D \rho}(k)$ might encode sufficiently identifiable signatures of the $D_1(2430)^0$ and $D_1(2420)$ states when smaller sources are considered.  It should be emphasized that if the femtoscopic analysis for the mentioned channels is done and the measured genuine CFs have similar behavior to those obtained here, then it is possible to say that this work provides a framework compatible with the existence of both broad and narrow states. Determining information on the $D^*\pi$ channel, where both mesons are not electrically charged, from smaller sources, like proton-proton collisions, can be useful in settling the value of the $D^*\pi$ scattering length. It should be possible to determine data on $D^{*0}\pi^+$ where $D^{*0}$ is reconstructed from $D^{+}\pi^-$.  
In this sense, it would be interesting to confront these CFs with data collected in future high-precision experiments.

\section{Conclusions}
The main conclusions of the discussions presented in this work can be summarized as follows:
\begin{itemize}
    \item The information on the properties of the two lightest $D_1$ states comes mostly from fits made to the experimental data on the $D^*\pi$ invariant mass, which gets contribution from several charm states. Such a procedure attributes similar masses but very different widths to the two of them. Although two states with similar masses are expected from the quark model, and once the masses are assumed one could explain their different widths on the basis of heavy quark symmetry, it seems not trivial to simultaneously describe both (masses and widths) from the same dynamics. Different models lead to different results and considerations of hadronic loops or similar mechanisms of mixing between $1^1P1$ and $1^3P1$  quark model seems to work better. 
    \item There exists information on the scattering length, determined in the lattice QCD calculations, for the $D^*\pi$ and $D\pi$ channels. Both values are very similar. Several model calculations constrain their parameters using the former information and determine the $D\pi$ scattering length in the infinite volume by using physical masses. It can be argued that the scattering lengths for $D\pi$ and $D^*\pi$ can be similar, which provides valuable information on the study of the $D^*\pi$ system. 
    \item A value for the $D^*\pi$ scattering length is also available from heavy ion collisions but it does not agree with those mentioned in the previous point.
    \item With the purpose of understanding the properties of the two  $D_1$ states and finding alternative ways to extract information on the $D^*\pi$ scattering length, we consider a model where different meson-meson interactions and a bare quark model pole constitute the lowest order amplitudes. Such amplitudes are used as kernels to solve the Bethe-Salpeter equation in a coupled channel approach. Consequently, a narrow pole is found to get generated by the hadron dynamics and is related to $D_1(2420)$. A broader pole is also found, whose properties match those of $D_1(2430)$ when both hadron dynamics and a bare quark-model pole are considered.
    \item To contemplate the two aforementioned disagreeing values of the scattering lengths, we present two models. The two differ in the parameters related to the bare quark-model pole. 
    \item Using such amplitudes we determine correlation functions and find that such information on the $D^{*0}\pi^+$ and $D^{*0}\rho^+$ channels, determined from smaller source sizes, can bring useful information on the subject. 
 \end{itemize}

\section{Acknowledgements}

This work is partly supported by the Brazilian agencies CNPq (L.M.A.: Grants No. 309950/2020-1, No. 400215/2022-5, No. 200567/2022-5, and No. 308299/2023-0, K.P.K.: Grants No. 407437/2023-1 and No. 306461/2023-4, A.M.T: Grant No. 304510/2023-8), FAPESP (K.P.K.: Grant No. 2022/08347-9; A. M. T.: Grant No. 2023/01182-7); and CNPq/FAPERJ under the Project INCT-F\'{\i}sica Nuclear e Aplicações (Contract No. 464898/2014-5). 
 
\bibliographystyle{apsrev4-1}
\bibliography{D1refs}
 
\end{document}